\documentclass[twocolumn,amsmath,amssymb,aps,nofootinbib]{revtex4-1}
\usepackage{bbm} 
\usepackage{eufrak} 
\usepackage{mathrsfs} 
\usepackage{tcolorbox}
\usepackage{ragged2e}
\usepackage{array,hhline,multirow}
\usepackage{floatrow,subfig,amssymb}
\usepackage[export]{adjustbox}
\usepackage{rotating} 
\usepackage{mathtools}
\allowdisplaybreaks[2]  
\usepackage{pifont,dsfont}
\usepackage{bm} 

\usepackage{graphicx,tikz}
\usepackage[framemethod=TikZ]{mdframed} 
\usepackage{tikz-cd} 
\usetikzlibrary{arrows,snakes,shapes.arrows,decorations.markings}
     \tikzset{>=triangle 90}
     \tikzstyle{bbc}=[draw,circle,fill=black,scale=.75]
     \tikzstyle{rc}=[circle,fill=red,scale=.6]
     \tikzstyle{wc}=[draw,circle,scale=.75]

\usepackage[colorlinks]{hyperref}
\hypersetup{
    colorlinks = true,
    linkcolor=blue,
    citecolor=magenta,
    linktocpage}

\def\bar{\overline}

\def\^{\wedge}

\def\m{{\mu}}

\def\ch{{\chi}}

\def\cA{{\mathcal A}}

\def\cC{{\mathcal C}}

\def\cD{{\mathcal D}}

\def\cH{{\mathcal H}}

\def\cI{{\mathcal I}}

\def\cL{{\mathcal L}}
\def\cM{{\mathcal M}}

\def\cN{{\mathcal N}}
\def\cO{{\mathcal O}}

\def\cT{{\mathcal T}}

\def\cV{{\mathcal V}}

\def\Mm{\cM_{M}[\cT]}

\newcommand{\cZ}{\mathcal{Z}}

\def\Csc{\mathscr{C}}

\def\H{\mathbb{H}}

\def\Z{\mathbb{Z}} 

\def\beq{\begin{equation}}
\def\eeq{\end{equation}}

\newcommand{\bpmat}{\begin{pmatrix}}
\newcommand{\epmat}{\end{pmatrix}}
\newcommand{\bsmat}{\begin{smallmatrix}}
\newcommand{\esmat}{\end{smallmatrix}}
\newfloatcommand{capbtabbox}{table}[][\FBwidth]

\def\bar{\overline}

\def\^{\wedge}

\def\Mm{\cM_{M}[\cT]}

\def\m{{\mu}}

\def\ch{{\chi}}

\def\cA{{\mathcal A}}

\def\cC{{\mathcal C}}

\def\cD{{\mathcal D}}

\def\cH{{\mathcal H}}

\def\cI{{\mathcal I}}

\def\cL{{\mathcal L}}
\def\cM{{\mathcal M}}

\def\cN{{\mathcal N}}
\def\cO{{\mathcal O}}

\def\cT{{\mathcal T}}

\def\cV{{\mathcal V}}

\def\Mm

\def\Csc{\mathscr{C}}

\def\H{\mathbb{H}}

\def\Z{\mathbb{Z}} 

\def\beq{\begin{equation}}
\def\eeq{\end{equation}}

\usepackage{xcolor}

\definecolor{dgreen}{rgb}{0, 0.55, 0}
\definecolor{dorange}{rgb}{0.8, 0.33, 0.0}
\definecolor{dorange}{rgb}{1.0, 0.5, 0.31}
\definecolor{dgrey}{rgb}{0.3,0.3,0.3}
\definecolor{grey}{rgb}{0.9,0.9,0.9}

\def\bM{\begin{matrix}}
\def\eM{\end{matrix}}
\def\bar{\overline}

\def\^{\wedge}

\def\cC{{\mathcal C}}

\def\H{\mathbbm{H}}
\def\cH{{\mathcal H}}

\def\cL{{\mathcal L}}
\def\cM{{\mathcal M}}

\def\cN{{\mathcal N}}
\def\cO{{\mathcal O}}

\def\cT{{\mathcal T}}

\def\cV{{\mathcal V}}

\def\Z{\mathbbm{Z}}

\def\m{{\mu}}

\def\ch{{\chi}}

\def\ZZ{{\mathbb{Z}}}
\def\cA{{\mathcal{A}}}
\def\cD{{\mathcal{D}}}
\def\cL{{\mathcal{L}}}

\def\half{{1 \over 2}}
\def\no{\nonumber}
\newcommand{\bea}{\begin{eqnarray}}
\newcommand{\eea}{\end{eqnarray}}

\usetikzlibrary{arrows.meta}

\newcommand{\opmidarrow}{\tikz[baseline] \draw[{Stealth[scale=1]}-] (0,0) -- +(.1,0);}

\usetikzlibrary{decorations.pathreplacing,calligraphy}
\usetikzlibrary{calc}

\tikzset{
  on each segment/.style={
    decorate,
    decoration={
      show path construction,
      moveto code={},
      lineto code={
        \path [#1]
        (\tikzinputsegmentfirst) -- (\tikzinputsegmentlast);
      },
      curveto code={
        \path [#1] (\tikzinputsegmentfirst)
        .. controls
        (\tikzinputsegmentsupporta) and (\tikzinputsegmentsupportb)
        ..
        (\tikzinputsegmentlast);
      },
      closepath code={
        \path [#1]
        (\tikzinputsegmentfirst) -- (\tikzinputsegmentlast);
      },
    },
  },
  mid arrow/.style={postaction={decorate,decoration={
        markings,
        mark=at position .5 with {\arrow[#1]{stealth}}
      }}},
}

\tikzset{line/.style={line width=0.25mm},
curve/.style={line,smooth,tension=1},
->-/.style={decoration={
  markings,
  mark=at position #1 with {\arrow[>={Stealth[scale=0.75]}]{>}}},postaction={decorate}},
-<-/.style={decoration={
  markings,
  mark=at position #1 with {\arrow[>={Stealth[scale=0.75]}]{<}}},postaction={decorate}},
}

\def\H{{\cH}}
\begin{document}

\begin{flushright}
\small KYUSHU-HET-343

\end{flushright}

\title{Haagerup Symmetry in $(E_8)_1$?}

\author{Jan Albert$^1$, Yamato Honda$^2$, Justin Kaidi$^{2,3,4}$, Yunqin Zheng$^5$}

\affiliation{$^1$Princeton Center for Theoretical Science, Princeton University, Princeton, NJ 08544, USA
\\
$^2$Department of Physics, Kyushu University, Fukuoka 819-0395, Japan
\\
$^3$Institute for Advanced Study, Kyushu University, Fukuoka 819-0395, Japan
\\
$^4$Quantum and Spacetime Research Institute (QuaSR), Kyushu University, Fukuoka 819-0395, Japan
\\
$^5$Kavli Institute for Theoretical Sciences,
University of Chinese Academy of Sciences, Beijing, 100190, China}

\date{\today}

\begin{abstract}
We suggest that the chiral $(\mathfrak{e}_8)_1$ theory---in many senses the simplest VOA---may have Haagerup symmetry $\H_i$ for $i=1,2,3$. Likewise, we suggest that the non-chiral $(E_8)_1$ WZW model may have $\H_i \times \H_i^\textrm{op}$ symmetry, and that gauging the diagonal symmetry gives a $c=8$ theory with $\cZ(\H_3)$ symmetry, which is the theory predicted in \cite{Evans:2010yr}. Along the way,  we  show that $(E_8)_1$ also has a $\mathrm{Fib} \times \mathrm{Fib}^\text{op}$ symmetry, and that gauging the diagonal symmetry gives the $(G_2)_1 \times (F_4)_1$ WZW model, explaining the well-known conformal embedding $(G_2)_1 \times (F_4)_1 \subset (E_8)_1$. Finally, we suggest a relation to theories with $\H_3$ symmetry at $c=2,6$, complimenting the discussion with new modular bootstrap results.

\end{abstract}

\maketitle 

Symmetry has long been a central organizing principle in theoretical physics, underlying phenomena from crystalline order to particle interactions. Traditional notions of symmetry are rooted in groups of transformations that act on physical systems, but recent advances in  quantum field theory and condensed matter physics have revealed a richer landscape of ``non-invertible'' symmetries, which cannot be described by ordinary groups, instead generalizing to algebraic structures known as fusion categories. Among these, the Haagerup categories \cite{haagerup1994principal,grossman2012quantum} stand out for being the simplest examples resisting realization in familiar quantum theories. In this work, we propose that the conformal field theory associated with the exceptional Lie algebra $E_8$ at level one---one of the simplest two-dimensional theories---naturally hosts Haagerup-type symmetry. These results provide a concrete new setting in which exotic symmetry structures may emerge in otherwise well-known quantum systems.


The Haagerup fusion categories $\H_i$ for $i=1,2,3$ where originally identified in \cite{haagerup1994principal,grossman2012quantum}, with the latter two having six simple objects and the following fusion ring, 
\bea
\label{eq:H3fusions}
&\vphantom{.}& \alpha^3 = 1~, \hspace{0.2 in} \alpha \rho = \rho \alpha^2~, 
\no\\
&\vphantom{.}& \rho^2 = 1 + \rho + \alpha \rho + \alpha^2 \rho~,
\eea
as is explained in more detail in the Supplemental Materials.
An explicit construction of a gapless theory with Haagerup symmetry was given in \cite{Bottini:2025hri}, though this theory has
multiple vacua \footnote{A gapped theory with Haagerup symmetry was constructed in \cite{Huang:2021ytb}. Non-unitary candidates for RCFTs with Haagerup symmetry have also appeared in \cite{Gang:2023ggt}.}.
On the other hand, a variety of recent results point to the existence of a gapless theory with a unique vacuum with Haagerup symmetry---for example, a series of numerical results have hinted at the existence of such a theory around central charge $c\approx 2$ \cite{Wolf:2020qdo,Huang:2021nvb,Vanhove:2021zop,Liu:2022qwn,Hung:2025gcp}.

In significantly older work, Evans and Gannon argued for the existence of $c\in8\ZZ$ theories with $\cZ(\H_3)$ symmetry, with $\cZ(\H_3)$ being the Drinfeld center of $\H_3$  \cite{Evans:2010yr}. Though they were unable to give a complete construction of such theories, they were able to use the modular data of $\cZ(\H_3)$ to constrain the modular characters, and in the case of $c=8$ they noticed an intriguing connection to the chiral $(\mathfrak{e}_8)_1$ theory, as well as its sub-VOA $(\mathfrak{e}_6)_1 \times (\mathfrak{a}_2)_1$.

In the current work, we offer an explanation for this connection at the level of the non-chiral $(E_8)_1$ WZW model; to avoid confusion, we will denote the chiral theory by $(\mathfrak{e}_8)_1$, and the non-chiral theory by $(E_8)_1$. In particular, we will suggest that $(E_8)_1$ has an $\H_3 \times \H_3^\text{op}$ symmetry \footnote{Here $\cC^\text{op}$ represents the opposite category to $\cC$, which has the same objects and fusion rules but with oppposite associators. For braided categories, the braiding is also inverted.}, and that gauging the diagonal $\cH_3$ symmetry (using recent techniques for non-invertible gauging developed in e.g. \cite{Frohlich:2009gb,Diatlyk:2023fwf,Perez-Lona:2023djo,Perez-Lona:2024sds}) gives the putative $\cZ(\H_3)$ theory predicted by Evans and Gannon. The role of the invertible $\ZZ_3$ element in the diagonal $\cH_3$ is played by the $\ZZ_{3\mathrm{B}}$ conjugacy class in $E_8$, whose centralizer is $E_6 \times A_2$.

We will also see that  $(\mathfrak{e}_8)_1$ is self-dual under certain gaugings, which would imply that $(\mathfrak{e}_8)_1$ has  $\H_i$ symmetry for $i=1,2,3$, as well as a number of additional duality defects. Finally, we will discuss connections to the $c\approx2$ theory predicted in \cite{Huang:2021nvb,Vanhove:2021zop,Liu:2022qwn,Hung:2025gcp}, suggest the existence of a $c\approx 6$ counterpart, and compliment the discussion with new modular bootstrap results, summarized in Figures \ref{fig:V1} and \ref{fig:V34}. Details will be given in upcoming work \cite{upcoming}.

Let us also note that Appendix \ref{app:lifts}, which gives a simple prescription for computing the change in fusion categories upon discrete gauging, may be of independent interest to the reader.

\section{Warm-Up: Automorphisms of $E_8$}
By far the most well-understood class of symmetries in two-dimensional rational conformal field theories (RCFTs) are those generated by Verlinde lines \cite{verlinde1988fusion,Petkova:2000ip}. Verlinde lines  commute not only with the Virasoro generators, but also with the full chiral algebra of the RCFT. It is the latter property which enables their straightforward classification, and for diagonal RCFTs they can be shown to be in one-to-one correspondence with RCFT primaries, further sharing the same fusion rules. 

However, Verlinde lines are not the only symmetries of RCFTs. Indeed, there can be a variety of line operators which are topological, but which do not commute with the chiral algebra. The current letter will focus on a certain class of such symmetries in the $(E_8)_1$  WZW model.\footnote{Previous discussions of symmetries of $(E_8)_1$ which do not commute with the chiral algebra can be found in \cite{Burbano:2021loy,Hegde:2021sdm,Rayhaun:2023pgc,Moller:2024xtt}. } 

The basic idea can be illustrated via the following simple example. Consider the well-known conformal embedding $(E_7)_1\times (A_1)_1 \subset (E_8)_1$. Unlike $(E_8)_1$, which has no non-trivial primary operators, and hence whose torus partition function can be written solely in terms of the vacuum character, 
\bea
\label{eq:E8partfunct}
Z_{E_8}(\tau) := |\chi^{E_8}_0|^2~, 
\eea
with
\bea
\label{eq:E8character}
\chi^{E_8}_0 = q^{-1/3} \left( 1 + 248 q + 4124 q^2 + 34752 q^3 + \dots \right)\,,\,\,\,
\eea
each of $(E_7)_1$ and $ (A_1)_1$ has a single non-trivial primary, with respective conformal weights $h= {3\over 4}$ and ${1\over 4}$. As such, the diagonal torus partition function is written as a sum of four terms, 
\bea
\label{eq:E7A1pf}
Z_{E_7 \times A_1}(\tau) &= &|\chi^{E_7}_0 \chi^{A_1}_0|^2 +|\chi^{E_7}_{3/4} \chi^{A_1}_0|^2
\no\\
&\vphantom{.}& \hspace{0.25 in} +|\chi^{E_7}_0 \chi^{A_1}_{1/4}|^2 +|\chi^{E_7}_{3/4} \chi^{A_1}_{1/4}|^2 ~. 
\eea
In this diagonal theory, there is a Verlinde line corresponding to each non-trivial primary, which in total generate a $\ZZ_2 \times \ZZ_2$ symmetry. 
In particular, we have topological lines $\eta_i$ for $i=1,2$ such that 
\bea
\eta_1 | 1\rangle^{E_7} &=& | 1\rangle^{E_7}~, \hspace{0.3 in}\eta_1 | 3/4 \rangle^{E_7} =-  | 3/4 \rangle^{E_7}~,
\no\\
\eta_2 | 1\rangle^{A_1} &=& | 1\rangle^{A_1}~, \hspace{0.3 in}\eta_2 | 1/4 \rangle^{A_1} =-  | 1/4 \rangle^{A_1}~,
\eea
and the remaining actions are trivial.

It is straightforward to verify that gauging the diagonal symmetry generated by $\eta := \eta_1 \eta_2$  takes us from the diagonal $(E_7)_1 \times (A_1)_1$ WZW model to the  $(E_8)_1$ WZW model. Indeed, computing the twisted-twined partition functions, 
\bea
Z_{E_7 \times A_1}|^\eta &= &|\chi^{E_7}_0 \chi^{A_1}_0|^2 -|\chi^{E_7}_{3/4} \chi^{A_1}_0|^2
\no\\
&\vphantom{.}& \hspace{0.25 in} -|\chi^{E_7}_0 \chi^{A_1}_{1/4}|^2 +|\chi^{E_7}_{3/4} \chi^{A_1}_{1/4}|^2 ~,
\no\\
Z_{E_7 \times A_1}|_\eta &= &\chi^{E_7}_0 \chi^{A_1}_{0}\overline \chi^{E_7}_{3/4} \overline \chi^{A_1}_{1/4}  
\no\\
&\vphantom{.}& \hspace{0.25 in} +\chi^{E_7}_0 \chi^{A_1}_{1/4}\overline \chi^{E_7}_{3/4} \overline \chi^{A_1}_{0} + \mathrm{c.c.}~,
\no\\
Z_{E_7 \times A_1}|^\eta_\eta &= &\chi^{E_7}_0 \chi^{A_1}_{0}\overline \chi^{E_7}_{3/4} \overline \chi^{A_1}_{1/4}  
\no\\
&\vphantom{.}& \hspace{0.25 in} -\chi^{E_7}_0 \chi^{A_1}_{1/4}\overline \chi^{E_7}_{3/4} \overline \chi^{A_1}_{0} + \mathrm{c.c.}~, 
\eea
 taking their sum, and dividing by the order of the group (two), we immediately land upon 
\bea
Z_{E_7 \times A_1/\eta}(\tau) = |\chi^{E_7}_0 \chi^{A_1}_{0} + \chi^{E_7}_{3/4} \chi^{A_1}_{1/4}|^2~,
\eea
which is equivalent to the $(E_8)_1$ partition function in (\ref{eq:E8partfunct}), upon recalling the well-known fact that
\bea
\chi^{E_8}_0 = \chi^{E_7}_0 \chi^{A_1}_{0} + \chi^{E_7}_{3/4} \chi^{A_1}_{1/4}~, 
\eea
which follows from the decomposition $\mathbf{248} \rightarrow (\mathbf{133}, \mathbf{1}) \oplus (\mathbf{1}, \mathbf{3}) \oplus (\mathbf{56},\mathbf{2})$.

Because it was possible to go from $(E_7)_1 \times (A_1)_1$ to  $(E_8)_1$ by gauging a discrete symmetry, there must exist a symmetry which can be gauged to go back. This symmetry will clearly not commute with the chiral algebra of $E_8$, and instead should be such that its centralizer in $E_8$ is $E_7 \times A_1$. Of course, the symmetry in this case is well-known---it is the subgroup $\ZZ_{2\mathrm{A}}\subset \mathrm{Aut}(E_8)$ (see e.g.~\cite{Kac:1996nq}). 

However, let us assume that we did not know this. A simpler way to arrive at this is to ``lift'' the Verlinde lines of $(E_7)_1 \times (A_1)_1$ in the following way (this lifting procedure is discussed in more detail in Appendix \ref{app:lifts}). Since $\eta$ is trivial in $(E_8)_1$, let us focus on $\eta_2$. We now use the action of $\eta_2$ on primaries in $(A_1)_1$ to define two operations in $(E_8)_1$, acting on the chiral and anti-chiral characters as 
\bea
\label{eq:etaLRdef}
\widehat \eta_L : \hspace{0.1 in} (\chi^{E_7}_0 \chi^{A_1}_{0},  \chi^{E_7}_{3/4} \chi^{A_1}_{1/4}) \rightarrow  (\chi^{E_7}_0 \chi^{A_1}_{0}, - \chi^{E_7}_{3/4} \chi^{A_1}_{1/4})~, ~~
\no\\
\widehat \eta_R : \hspace{0.1 in} (\overline \chi^{E_7}_0 \overline\chi^{A_1}_{0},  \overline\chi^{E_7}_{3/4} \overline\chi^{A_1}_{1/4}) \rightarrow  (\overline\chi^{E_7}_0 \overline\chi^{A_1}_{0},  -\overline\chi^{E_7}_{3/4} \overline\chi^{A_1}_{1/4})~.~~
\eea
Each of these can be thought of as defining a symmetry in one of the two chiral halves $(\mathfrak{e}_8)_1$ making up the $(E_8)_1$ theory. 

While both of these symmetries are anomalous (see e.g.\ Table 2 in \cite{Burbano:2021loy}), the diagonal symmetry $\widehat \eta := \widehat \eta_L \widehat \eta_R$ is non-anomalous, and gauging it takes us back to $(E_7)_1 \times (A_1)_1$---in other words, $\widehat \eta$ generates $\ZZ_{2\mathrm{A}}$. This can be seen by computing the twisted-twined partition functions via (\ref{eq:etaLRdef}) and the modular $S$-matrix, giving
\bea
Z_{E_8}(\tau)|^{\widehat\eta} &=& |\chi^{E_7}_0 \chi^{A_1}_{0} - \chi^{E_7}_{3/4} \chi^{A_1}_{1/4}|^2~,
\no\\
Z_{E_8}(\tau)|_{\widehat\eta} &=& |\chi^{E_7}_{3/4} \chi^{A_1}_{0} + \chi^{E_7}_{0} \chi^{A_1}_{1/4}|^2~,
\no\\
Z_{E_8}(\tau)|^{\widehat\eta}_{\widehat\eta} &=& |\chi^{E_7}_{3/4} \chi^{A_1}_{0} - \chi^{E_7}_{0} \chi^{A_1}_{1/4}|^2~,
\eea
and taking the relevant sum to reproduce (\ref{eq:E7A1pf}). 

Of course, the exact same techniques can be used more generally. For example, gauging the diagonal $\ZZ_3$ center symmetry of $(E_6)_1 \times (A_2)_1$ gives rise again to $(E_8)_1$, and one can go back by gauging $\ZZ_{3 \mathrm{B}}\subset\mathrm{Aut}(E_8)$, which does not commute with the chiral algebra. The action of $\ZZ_{3 \mathrm{B}}$ can be defined as above, by first decomposing the characters of $(E_8)_1$ into those of $(E_6)_1 \times (A_2)_1$, 
\bea
\chi^{E_8}_0 = \chi^{E_6}_0 \chi^{A_2}_{0} + \chi^{E_6}_{2/3 } \chi^{A_2}_{1/3 }+ \chi^{E_6}_{\bar{2/3} } \chi^{A_2}_{\bar{1/3} }~,
\eea
and then ``lifting'' the action of the center symmetry of one of the two factors to $(\mathfrak{e}_8)_1$. 

In both of the cases mentioned above, the symmetry in question was an element of $\mathrm{Aut}(E_8)$, and hence its existence was no surprise. 
A slightly more interesting example is to consider the conformal embedding $(G_2)_1 \times (F_4)_1 \subset (E_8)_1$. As described in Appendix \ref{app:Fib}, each of $(G_2)_1$ and $(F_4)_1$ has a Fibonacci symmetry, and we may gauge the diagonal symmetry to obtain $(E_8)_1$. Conversely,  the Verlinde lines of $(G_2)_1 \times (F_4)_1$ may be lifted to a $\mathrm{Fib} \times \mathrm{Fib}^\mathrm{op}$ symmetry of $(E_8)_1$, and gauging the diagonal symmetry takes us back to $(G_2)_1 \times (F_4)_1$.

\section{Haagerup Symmetry}

Having introduced the basic idea, we may now move on to the case of actual interest in this letter. The starting point is the result of \cite{Evans:2010yr,gannon2019reconstruction}, which claims that any vector-valued modular form whose modular data is that of the Drinfeld center $\cZ(\H_3)$ and has central charge $c=8$ must have characters of the following form, 
\begin{equation}
    \label{eq:ZH3characters}
    \begin{split}
\chi_0^{\cZ(\H_3)} &=  q^{2/3} \left(q^{-1} + (6 + 13 \gamma) + (120 + 78 \gamma) q \right.
\\
&\vphantom{.} \hspace{0.1 in}\left.+ (956 + 351 \gamma) q^2 + (6010 + 1235 \gamma) q^3 + \dots \right)~,
\\
\chi_{\pi_1}^{\cZ(\H_3)}   &=  q^{2/3} \left( (80-13 \gamma) + (1250 -78 \gamma) q \vphantom{q^3}\right.
\\
&\vphantom{.} \hspace{0.1 in}\left.+ (10630 - 351 \gamma) q^2 +(65042-1235\gamma) q^3 +  \dots \right)~,
\\
\chi_{\pi_2}^{\cZ(\H_3)}   &= \chi_{\sigma_1}^{\cZ(\H_3)}  \\
&= q^{2/3} \left( 81 + 1377 q + 11583 q^2 + 71037 q^3+ \dots \right)~,
\end{split}
\end{equation}
where $\gamma = 0$ or $1$ and the characters are labelled by elements of $\cZ(\H_3)$. Note that we have  written only  four of the twelve total characters; the others can be found in \cite{Evans:2010yr}. We refer to the putative diagonal, non-chiral CFT with these characters as the $\cZ(\H_3)$ theory.\footnote{We will assume that this $c=8$ theory actually exists---to prove that $(E_8)_1$ actually has Haagerup symmetry, it would be necessary to prove the existence of this theory. However, let us note that our computations relating $\cZ(\H_3)$ and $\H_3 \times \H_3^\mathrm{op}$ symmetries hold more generally, relating any theory with  $\cZ(\H_3)$ symmetry to one with $\H_3 \times \H_3^\mathrm{op}$ symmetry. }

The key observation is that we have the following relation between characters
\bea
\label{eq:E8Haagrel}
\chi_0^{E_8} &=&  \chi_0^{\cZ(\H_3)}   + \chi_{\pi_1}^{\cZ(\H_3)} + 2\, \chi_{\pi_2}^{\cZ(\H_3)}~,
\no\\
&=&  \chi_0^{\cZ(\H_3)}   + \chi_{\pi_1}^{\cZ(\H_3)} + 2\, \chi_{\sigma_1}^{\cZ(\H_3)}~,
\no\\
&=&  \chi_0^{\cZ(\H_3)}   + \chi_{\pi_1}^{\cZ(\H_3)} +  \chi_{\pi_2}^{\cZ(\H_3)}+\chi_{\sigma_1}^{\cZ(\H_3)}~.
\eea
Focusing for the moment on the first of these, this relationship suggests that the $(E_8)_1$ WZW model can be obtained by gauging the algebra object $\cA = 1 \oplus \pi_1 \oplus 2 \pi_2$ in the putative $\cZ(\H_3)$ theory. Upon gauging this algebra object, a standard module category analysis shows that the resulting theory has symmetry $\H_3 \times \H_3^{\mathrm{op}}$, which leads us to the claim that  $(E_8)_1$ has $\H_3 \times \H_3^{\mathrm{op}}$ symmetry.\footnote{A similar suggestion has been made previously in \cite{Moller:2024xtt}. } This can also be  shown by lifting the Verlinde lines of $\cZ(\H_3)$ to $(E_8)_1$ via the same procedure as for $(E_7)_1 \times (A_1)_1 \subset (E_8)_1$, as is discussed in Appendix \ref{app:E8lifts}. 
Here in the main text, we will content ourselves with briefly describing the action of this symmetry, and confirming that gauging of the diagonal Haagerup symmetry gives the $\cZ(\H_3)$ theory of \cite{Evans:2010yr}.\footnote{As a more general fact, note that for any fusion category $\cC$, the Drinfeld center $\cZ(\cC)$ and $\cC \times \cC^\mathrm{op}$ are Morita equivalent---$\cZ(\cC)$ can be obtained from $\cC \times \cC^\mathrm{op}$ by gauging the Frobenius algebra $\mathcal{A}= \oplus_{c\in \cC} c\boxtimes c^{\mathrm{op}}$ \cite{Mueger:2001crc}. }

The idea is similar to before---namely, each copy of $\H_3$ acts on a single chiral half, with the action given by the usual action of $\H_3$ on elements of $\cZ(\H_3)$. Focusing on  the left-movers, we have \cite{Lin:2022dhv,Bottini:2025hri} (see also Appendix \ref{app:review}), 
\bea
&\vphantom{.}& \chi_0^{\cZ(\H_3)}: \hspace{0.1in} \widehat \alpha_L = 1~,\,\, \widehat \rho_L = d_\rho~, 
\no\\
&\vphantom{.}&\chi_{\pi_1}^{\cZ(\H_3)}: \hspace{0.1in} \widehat \alpha_L = 1~,\,\, \widehat \rho_L = -d_\rho^{-1}~,
\no\\
&\vphantom{.}& \chi_{\pi_2}^{\cZ(\H_3)}: \hspace{0.1 in} \widehat \alpha_L = \left(\begin{smallmatrix} \omega & 0 \\ 0 & \omega^2 \end{smallmatrix} \right)~, \hspace{0.1in} \widehat \rho_L = \left(\begin{smallmatrix}0 & 1 \\ 1 & 0  \end{smallmatrix} \right) ~,
\eea
where $d_\rho := {1\over 2}(3 + \sqrt{13})$ and $\omega:=e^{2 \pi i /3}$, and likewise for the right-movers. Using these results and (\ref{eq:E8Haagrel}), we then arrive at the following partition functions with insertions of the diagonal operators $\widehat \alpha : = \widehat \alpha_L \widehat \alpha_R^{-1}$ and $\widehat \rho :=\widehat  \rho_L\widehat \rho_R$,
\bea
\label{eq:alpharhoinpf}
Z_{E_8}|^{\widehat \alpha} &=& \left|\chi_0^{\cZ(\H_3)} + \chi_{\pi_1}^{\cZ(\H_3)} + (\omega + \omega^2) \chi_{\pi_2}^{\cZ(\H_3)} \right|^2~, 
\no\\
Z_{E_8}|^{\widehat \rho} &=& \left|d_\rho \chi_0^{\cZ(\H_3)} - d_\rho^{-1} \chi_{\pi_1}^{\cZ(\H_3)} \right|^2~.
\eea
Making use of the $S$-matrix for $\cZ(\H_3)$ reviewed in Appendix \ref{app:review}, we further obtain the twisted partition functions,
\begin{equation}
\begin{split}
\label{eq:twistedpartfuncts}
Z_{E_8}|_{\widehat \alpha} &= \left|\ch^{\cZ(\H_3)}_{\sigma_1} + \chi^{\cZ(\H_3)}_{\sigma_2} + \chi^{\cZ(\H_3)}_{\sigma_3} \right|^2~, 
\\
Z_{E_8}|_{\widehat \rho} &= \left| \chi^{\cZ(\H_3)}_{\pi_1} + \chi^{\cZ(\H_3)}_{\pi_2} + \sum_{i=1}^3 \chi^{\cZ(\H_3)}_{\sigma_i} + \sum_{i=1}^6 \chi^{\cZ(\H_3)}_{\mu_i} \right|^2~.
\end{split}
\end{equation}
Of course, similar results hold for $\widehat \alpha^2, \widehat \alpha \widehat \rho,$ and  $\widehat \alpha^2 \widehat \rho$.

We now claim that gauging the diagonal algebra $\cA^{\mathrm{diag}} := 1 \oplus \widehat \alpha \oplus \widehat \alpha^2 \oplus \widehat \rho \oplus\widehat \alpha \widehat \rho\oplus \widehat \alpha^2 \widehat \rho$ gives the $\cZ(\H_3)$ theory, with diagonal modular invariant
\bea\label{eq:diagonalZH3}
Z_{\cZ(\H_3)} &=& |\chi_0^{\cZ(\H_3)}|^2 + |\chi_{\pi_1}^{\cZ(\H_3)}|^2 +  |\chi_{\pi_2}^{\cZ(\H_3)}|^2
\no\\
&\vphantom{,}& \hspace{0.2 in} +\sum_{i=1}^3 |\chi^{\cZ(\H_3)}_{\sigma_i}|^2 + \sum_{i=1}^6 |\chi^{\cZ(\H_3)}_{\mu_i}|^2~.
\eea
The (co)multiplication for the algebra object may be taken to be the most obvious one, where all coefficients in the expansion in terms of simple lines are fixed to the same value, namely the inverse square root of the order of the algebra, i.e. $(3 + 3 d_\rho^2)^{-1/2}$. Thanks to the transparent nature of the Haagerup fusion category \cite{Huang:2020lox}, the gauging may be simplified significantly, and we are ultimately left to compute the following, 
\begin{equation}
\begin{split}
\label{eq:gaugedE8pf}
Z_{E_8 / \cA^{\mathrm{diag}} } = {1\over 3 + 3 d_\rho^2} \Bigg[ Z_{E_8} + 2 \left(Z_{E_8}|^{\widehat \alpha}+Z_{E_8}|_{\widehat \alpha} +Z_{E_8}|_{\widehat \alpha}^{\widehat \alpha}\right) 
\\
\left.+ 3 Z_{E_8}|_{\widehat \rho,1}^{\widehat \rho} + 3 \sum_{i=0}^2 \left( Z_{E_8}|_{\widehat \rho,\widehat \alpha^i \widehat \rho }^{\widehat \rho} +Z_{E_8}|_{\widehat \rho,\widehat \alpha^i \widehat \rho }^{\widehat\alpha \widehat \rho} +Z_{E_8}|_{\widehat \rho,\widehat \alpha^i \widehat \rho }^{\widehat\alpha^2 \widehat \rho}\right)  \right]~,
\end{split}
\end{equation}
where
\bea
Z|^x_{y,z} \,\,\,\,&:=&\,\,\,\, \begin{tikzpicture}[baseline={([yshift=+.5ex]current bounding box.center)},vertex/.style={anchor=base,
    circle,fill=black!25,minimum size=18pt,inner sep=2pt},scale=0.4]
    \filldraw[grey] (-2,-2) rectangle ++(4,4);
    \draw[thick, dgrey] (-2,-2) -- (-2,+2);
    \draw[thick, dgrey] (-2,-2) -- (+2,-2);
    \draw[thick, dgrey] (+2,+2) -- (+2,-2);
    \draw[thick, dgrey] (+2,+2) -- (-2,+2);
    \draw[thick, black, -stealth] (0,-2) -- (0.354,-1.354);
    \draw[thick, black] (0,-2) -- (0.707,-0.707);
    \draw[thick, black, -stealth] (2,0) -- (1.354,-0.354);
    \draw[thick, black] (2,0) -- (0.707,-0.707);
    \draw[thick, black, -stealth] (-0.707,0.707) -- (-0.354,1.354);
    \draw[thick, black] (0,2) -- (-0.707,0.707);
    \draw[thick, black, -stealth] (-0.707,0.707) -- (-1.354,0.354);
    \draw[thick, black] (-2,0) -- (-0.707,0.707);
    \draw[thick, black, -stealth] (0.707,-0.707) -- (0,0);
    \draw[thick, black] (0.707,-0.707) -- (-0.707,0.707);

    \node[black, below] at (0,-2) {\scriptsize $y$};
    \node[black, right] at (2,0) {\scriptsize $x$};
    \node[black, above] at (0.2,0) {\scriptsize $z$};
\end{tikzpicture} ~.
\eea

The terms in the first line, together with the first term on the second line, are all straightforward to compute. The only non-trivial contributions are from the terms of the form $ Z_{E_8}|_{\widehat \rho,\widehat \alpha^i \widehat \rho }^{\widehat \alpha^j \widehat \rho} $. These may be computed by explicitly solving the tube algebra, which in particular requires making use of the $F$-symbols given in \cite{Huang:2020lox,Huang:2021ytb}; details can be found in Appendix \ref{app:review}. Doing so tells us, for example, that we have
\bea
 Z_{E_8}|_{\widehat \rho,\widehat \rho }^{\widehat \rho}  &\approx& \big|-0.128\, \chi_{\pi_1} + 0.845\, \chi_{\pi_2} + 0.422\, \chi_{\sigma_1}
 \no\\
 &\vphantom{.}&\hspace{0.3 in} +\, 1.39\, ( e^{- 2 \pi i {1\over 3}}\chi_{\sigma_2} +e^{2 \pi i {1\over 3}} \chi_{\sigma_3}) 
 \\
&\vphantom{.}& \hspace{0.3 in} +\, 0.509( e^{2 \pi i {1\over 13}} \chi_{\mu_1} + e^{-2 \pi i {1\over 13}} \chi_{\mu_2}) 
 \no\\
  &\vphantom{.}&\hspace{0.3 in}  +\,  0.296( e^{2 \pi i {9\over 13}} \chi_{\mu_3} + e^{-2 \pi i {9\over 13}} \chi_{\mu_4}) 
  \no\\
 &\vphantom{.}&\hspace{0.3 in}  -\,1.65 ( e^{2 \pi i {3\over 13}} \chi_{\mu_5} + e^{-2 \pi i {3\over 13}} \chi_{\mu_6})  \big|^2~, 
 \no
\eea
where we have written only numerical approximations of the coefficients, since their closed form expressions are complicated and unenlightening. The other partition functions are obtained similarly.

Computing all of the relevant twisted partition functions and inserting them into (\ref{eq:gaugedE8pf}), one straightforwardly confirms that $Z_{E_8 / \mathcal{A}^\mathrm{diag}}  = Z_{\cZ(\H_3)} $, as claimed. Note that the $\widehat \alpha$ appearing in $\cA^\mathrm{diag}$ is none other than the generator of $\ZZ_{3 \mathrm{B}} \subset \mathrm{Aut}(E_8)$, as can be seen from the fact that 
\bea
&\vphantom{.}& \chi_0^{E_6} \chi_{0}^{A_2} = \chi_0^{\cZ(\H_3)} + \chi_{\pi_1}^{\cZ(\H_3)} ~, 
\no\\
&\vphantom{.}& \chi_{2/3}^{E_6} \chi_{1/3}^{A_2} +\chi_{\overline{2/3}}^{E_6}\chi_{\overline{1/3}}^{A_2}= 2 \chi^{\cZ(\H_3)}_{\pi_2} ~,
\eea
together with the action in (\ref{eq:alpharhoinpf}). 

Returning now to the other equalities in (\ref{eq:E8Haagrel}), it is easy to verify that one may go between the first and second lines by gauging the chiral $\widehat \alpha_L$, which tells us that $(\mathfrak{e}_8)_1$ theory should also have $\H_2 $ symmetry, since $\H_2$ and $\H_3$ are related by $\ZZ_3$ gauging \cite{Huang:2021ytb}. Likewise, the third equality is obtained by gauging $\cA_{L} = 1 \oplus \widehat \rho_{L} \oplus\widehat \alpha_{L} \widehat \rho_{L} $, from which we learn that $(\mathfrak{e}_8)_1$ should also have $\H_1$ symmetry. Additional details can be found in Appendix \ref{app.chiral}. Similar statements hold for the anti-chiral halves.

Let us close by noting that by a similar reasoning, the Monster CFT is also expected to have $\H_i \times \H_i^\mathrm{op}$ symmetry for $i=1,2,3$, though we do not give details here.

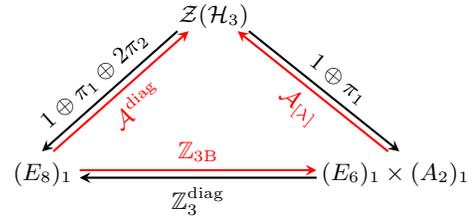
\begin{figure}[!t]
\begin{tikzpicture}

     \node[right]  at (-2.3,0) {$(E_8)_1$};
     \node[left] at (4,0) {$(E_6)_1 \times (A_2)_1  $};
      \node[right]  at (-0.1,2.1) {$\cZ(\H_3) $};

    \draw[-stealth,thick] (1.85,-0.05)--(-1.3,-0.05) node[midway,  below] {$\ZZ_3^{\mathrm{diag}}$};
   \draw[-stealth,thick ,red] (-1.3,0.05)--(1.85,0.05) node[midway,  above,red] {$\ZZ_{3\mathrm{B}}$};
        
  \draw[-stealth,thick,red] (-1.7+0.05,0.3)--(0.2+0.05-0.1,1.9) node[midway,  red,sloped, below] {$\,\mathcal{A}^\mathrm{diag}$};
  \draw[-stealth,thick] (0.2-0.1-0.1,1.9)--(-1.7-0.1,0.3) node[midway,  sloped,above] {$1 \oplus \pi_1 \oplus 2 \pi_2\,\,$};
  
    \draw[-stealth,thick,red] (3-0.1-0.1,0.3)--(1-0.1-0.1,1.9) node[midway,  red,sloped,below] {$\cA_{[ \lambda]}\,$};
  \draw[-stealth,thick] (1+0.05-0.1,1.9)--(3+0.05-0.1,0.3) node[midway,  sloped,above] {$\,\,1 \oplus \pi_1$};

    \end{tikzpicture}
    \caption{Gaugings relating the $\cZ(\H_3)$, $(E_8)_1$, and $(E_6)_1 \times (A_2)_1$ theories. Lines in red denote symmetries which do not commute with the full chiral algebra. }
\label{fig:gaugings}
\end{figure}

\begin{figure}[!tbp]
\begin{center}
\includegraphics[width=0.8\linewidth]{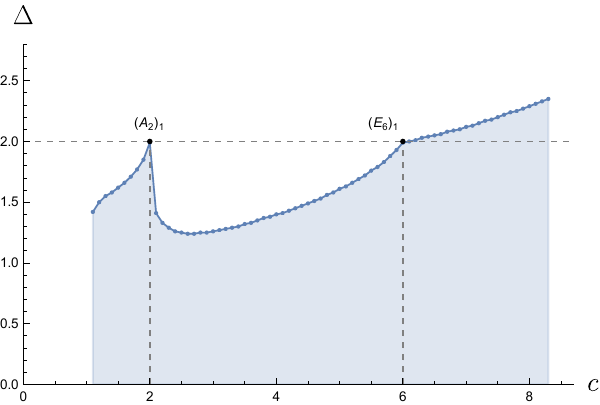}
\caption{Bounds on the dimension $\Delta$ of the lightest scalar in $\cV_1$, as a function of $c$. The allowed region is shaded in blue. Note that both the $(A_2)_1$ and $(E_6)_1$ theories live at kinks on the boundary.
}
\label{fig:V1}
\end{center}
\end{figure}

\section{Relation to $c=2,6$}

As we have described above, we may gauge the $\ZZ_{3\mathrm{B}}$ symmetry of $(E_8)_1$ theory to go to the $(E_6)_1 \times (A_2)_1$ WZW model. As such, the latter should contain a symmetry $(\H_3 \times \H_3^\mathrm{op}) / \ZZ_{3\mathrm{B}}$, where $\ZZ_{3\mathrm{B}}$ is the diagonal symmetry of the two copies of $\H_3$. As we discuss in Appendix \ref{app:E6A2lifts}, this category is equivalent to a double near-group extension of $\ZZ_3^2$. Choosing one of the dimension  $d_\rho^2$ lines in this category, which we denote by $[ \lambda]$, we may then gauge the binary algebra $\cA_{[ \lambda]}:= 1 \oplus[ \lambda]$, which takes us to the $\cZ(\H_3)$ theory. Conversely, gauging the object $1 \oplus \pi_1$ in the  $\cZ(\H_3)$ takes us back to $(E_6)_1 \times (A_2)_1$. These relations are summarized in Figure \ref{fig:gaugings}.

\begin{figure}[!tbp]
\begin{center}
\includegraphics[width=0.8\linewidth]{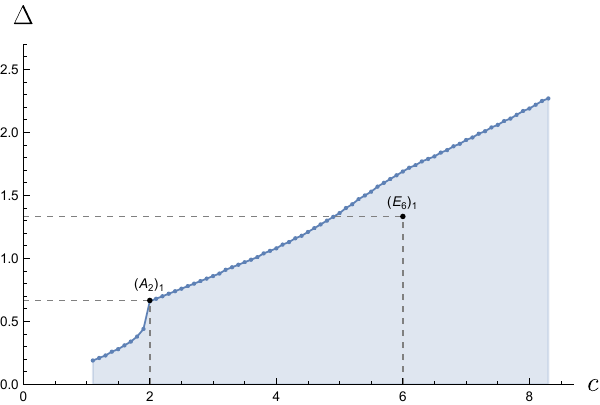}
\caption{Bounds on the dimension $\Delta$ of the lightest $\cV_{\pi_2} \cup \cV_{\sigma_1}$ (i.e. $\ZZ_3$-charged) scalar, as a function of $c$. 
The $(A_2)_1$ theory again lives at a kink.  }
\label{fig:V34}
\end{center}
\end{figure}

Based on this, one may hope that upon performing appropriate (asymmetric) quotients, the above will give rise to theories with Haagerup symmetry at $c=2 ,6$. Details will be given in upcoming work \cite{upcoming}. Here, we offer just one tantalizing piece of evidence that this hope is correct---namely, we have performed numerical modular bootstrap for theories with $\mathcal H_i$ symmetry following the procedure in \cite{Lin:2019kpn,Lin:2023uvm}. The technique is described in Appendix \ref{app:modularbootstrap}, with the main results given in Figures \ref{fig:V1} and \ref{fig:V34}. The first of these gives bounds on the dimension of the lightest scalar operator in the identity sector, showing prominent kinks at $c\approx 2,6$ and $\Delta\approx 2$. The second gives bounds on the dimension of the lightest scalar operator in the $\pi_2 \cup \sigma_1$-sector, showing a kink at $c=2$ and $\Delta \approx {2\over 3}$.

\section*{Acknowledgements}

The numerical computations in this work were carried out on the Genkai cluster at Kyushu University and the Princeton Della cluster.
The authors thank Yuji Tachikawa for comments on a draft.  JA is grateful to Rajeev Erramilli and Justin Kulp for useful discussions, and acknowledges partial support from Simons Foundation Grant No.\ 917464.
YH is supported by Grant-in-Aid for JSPS Fellows No. 25KJ1925.
The research of JK is supported in part by the Inamori Foundation through the Inamori Frontier Program at Kyushu University. Y.Z. is supported by NSFC grant No.12505093 and the starting funds from University of Chinese Academy of Sciences (UCAS) and from the Kavli Institute for Theoretical Sciences (KITS).

\begin{center} \textit{This letter is written in loving memory of Thelma-Louise Kaidi. }
\end{center}

\bibliography{bib}

\appendix

\section{The category $\H_3$ and its Drinfeld center}
\label{app:review}

The Haagerup fusion category $\H_3$ has six simple objects 
$\{1, \alpha, \alpha^2, \rho, \alpha \rho, \alpha^2 \rho\}$ with fusion rules given in the main text, from which we read off the quantum dimensions 
$d_1 = d_\alpha = d_{\alpha^2}=1$ and $d_\rho = d_{\alpha \rho} = d_{\alpha^2 \rho} = {1 \over 2}(3 + \sqrt{13})$. 
Gauging the algebra object $\cA = 1 \oplus \alpha \oplus \alpha^2$ maps to a Morita equivalent fusion category denoted by $\H_2$, 
while gauging the object $\cA = 1 \oplus \rho \oplus \alpha \rho$ maps to a category denoted by $\H_1$ (see e.g. \cite[Section 8]{Huang:2021ytb} and Figure \ref{fig:gaugings2} below).

In order to solve the tube algebra for $\H_3$, we will need the $F$-symbols, which are given in unitary gauge in \cite{Huang:2020lox,Huang:2021ytb}. 
We take the convention in which the $F$-symbols are defined by 
 \bea
   \begin{tikzpicture}[baseline={([yshift=-1ex]current bounding box.center)},vertex/.style={anchor=base,
    circle,fill=black!25,minimum size=18pt,inner sep=2pt},scale=0.3]
   \draw[->-=0.2,->-=0.6,->-=0.95,thick] (0,0) -- (4,4);
   \draw[->-=0.7,thick] (2.6,0) -- (1.3,1.3);
    \draw[->-=0.6,thick] (6,0) -- (3,3);
          \node[below] at (0,-0.2) {$a$};
     \node[below] at (3,0) {$b$};
      \node[below] at (6,-0.2) {$c$};
     \node[above] at (4,4) {$d$};
      \node[right] at (2,2) {$e$};
    \end{tikzpicture} 
    = 
    \sum_{f} (F_{abc}^d)_{ef}\,\,\,
       \begin{tikzpicture}[baseline={([yshift=-1ex]current bounding box.center)},vertex/.style={anchor=base,
    circle,fill=black!25,minimum size=18pt,inner sep=2pt},scale=0.3]
   \draw[->-=0.45,->-=0.95,thick] (0,0) -- (4,4);
   \draw[->-=0.6,thick] (3,0) -- (4.5,1.5);
    \draw[->-=0.3,->-=0.8,thick] (6,0) -- (3,3);
          \node[below] at (0,-0.2) {$a$};
     \node[below] at (3,0) {$b$};
      \node[below] at (6,-0.2) {$c$};
     \node[above] at (4,4) {$d$};
      \node[right] at (2.5,1.8) {$f$};

    \end{tikzpicture} ~.
\eea
By using the transparency of the Haagerup fusion category, one can reduce all non-trivial $F$-symbols to those of the form  $(F^\rho_{\rho\rho\rho})_{11} = d_\rho^{-1}$ and  $(F^\rho_{\rho\rho\rho})_{1 \rho} =  d_{\rho}^{-1/2}$, as well as the cases  $(F^*_{\rho\rho\rho})_{* \rho}$ with both of the stars running over non-invertible elements. These are collected in the following table,  
\bea
\begin{tabular}{c|ccc}
$(F_{\rho\rho\rho}^*)_{*\rho}$ & $\rho$ & $\alpha \rho$ & $\alpha^2 \rho $
\\\hline 
$\rho$ & $x$ & $y_1$ & $y_2$ 
\\
$\alpha\rho$ & $y_1$ & $y_2$ & $z$ 
\\
$\alpha^2\rho$ & $y_2$ & $z$ & $y_1$ 
\end{tabular}~,
\eea
where we have defined the following constants, 
\bea
x &=& {1\over 3} (2- \sqrt{13})~, \hspace{0.4 in}z = {1\over 6} (1+ \sqrt{13})~,
\no\\
y_{1,2} &=& {1\over 12} \left( 5 - \sqrt{13} \mp \sqrt{6(1+ \sqrt{13})}\right) ~. 
\eea

We will also need the data of the Drinfeld center $\cZ(\H_3)$. There are twelve simple objects in this category, 
\bea
\{ 1, \pi_1, \pi_2, \sigma_1, \sigma_2, \sigma_3, \mu_1, \mu_2, \mu_3, \mu_4, \mu_5, \mu_6\}~,
\eea
whose quantum dimensions are given by,
\bea
d_{\pi_1} = 3d_\rho + 1~, \hspace{0.1 in} d_{\pi_2} = d_{\sigma_i} = 3d_\rho + 2~, \hspace{0.1 in} d_{\mu_i} = 3 d_\rho~.\no
\eea
The corresponding spins are, in order,
\bea
&\vphantom{,}& \{1,1,1,1, e^{2 \pi i \over 3},e^{-2 \pi i \over 3}, e^{4 \pi i \over 13},e^{-4 \pi i \over 13},
\no\\
&\vphantom{,}& \hspace{1 in} e^{10 \pi i \over 13},e^{-10 \pi i \over 13},e^{12 \pi i \over 13},e^{-12 \pi i \over 13}  \} ~.
\eea
The $S$-matrix is given as follows,\footnote{The modular data for $\cZ(\H_3)$ was first computed in \cite{Izumi:2001mi}, and later simplified in \cite{evans2006modular,Evans:2010yr}. We follow the presentation in \cite{Bottini:2025hri}.} 
\bea
\label{eq:ZH3Smatrix}
S = {1\over D} \left(\begin{matrix}  A & B \\ B^T & C\end{matrix} \right) ~,
\eea
where $D = {3\over 2} (13 + 3 \sqrt{13})$ and we have matrices
\bea
A &=& \left( \begin{smallmatrix}  
1 & 3 d_\rho + 1 &3 d_\rho + 2 & 3 d_\rho + 2 & 3 d_\rho + 2 & 3 d_\rho + 2
\\
 3 d_\rho + 1 & 1&  3 d_\rho + 2 & 3 d_\rho + 2 & 3 d_\rho + 2 & 3 d_\rho + 2
 \\
 3 d_\rho + 2 & 3 d_\rho + 2 & 6 d_\rho + 4 & -3 d_\rho - 2 & -3 d_\rho - 2 &  -3 d_\rho - 2
 \\
  3 d_\rho + 2 & 3 d_\rho + 2 & -3 d_\rho - 2  & 6 d_\rho + 4 & -3 d_\rho - 2 &  -3 d_\rho - 2
  \\
   3 d_\rho + 2 & 3 d_\rho + 2 & -3 d_\rho - 2  &  -3 d_\rho - 2 &  -3 d_\rho - 2 & 6 d_\rho + 4 
     \\
   3 d_\rho + 2 & 3 d_\rho + 2 & -3 d_\rho - 2  &  -3 d_\rho - 2 &  6 d_\rho + 4 &  -3 d_\rho - 2 
 \end{smallmatrix} \right) ~,
\no\\
B&=& 3 d_\rho \left(\begin{matrix} 1 & 1 & 1 & 1 & 1 & 1 
\\ 
-1& -1 & -1 & -1& -1&-1
\\
0 & 0 & 0 & 0 & 0 & 0 
\\ 
0 & 0 & 0 & 0 & 0 & 0 
\\
0 & 0 & 0 & 0 & 0 & 0 
\\
0 & 0 & 0 & 0 & 0 & 0  \end{matrix}  \right) ~,
\no\\
C &=& -6 d_\rho \left(\begin{matrix} 
c_4 & c_6 & c_1 & c_5 & c_3 & c_2
\\ 
c_6 & c_4 & c_5 &  c_1 & c_2 & c_3
\\
c_1 & c_5  & c_3 & c_2 & c_4 & c_6
\\ 
c_5 & c_1 & c_2 & c_3 & c_6 & c_4
\\
c_3 & c_2 & c_4 & c_6 & c_1 & c_5
\\
c_2 & c_3 & c_6 & c_4 & c_5 & c_1  \end{matrix}  \right) ~,
\eea
where $c_j:=\cos(\frac{2\pi j}{13})$.

There are three Lagrangian algebras in $\cZ(\H_3)$, namely 
\begin{equation}\label{eq:L123}
    \begin{split}
         \cL_1&= 1 \oplus \pi_1 \oplus \pi_2 \oplus \sigma_1~,\\
        \cL_2&=1 \oplus \pi_1 \oplus 2\sigma_1~,\\
        \cL_3&=1 \oplus \pi_1 \oplus 2\pi_2~.
    \end{split}
\end{equation}
Condensing any of these gives, in our context, the $(E_8)_1$ theory in one of the three presentations of (13) of the main text. 
The only other non-trivial algebra is $1 \oplus \pi_1$, which can be condensed to give a theory with reduced topological order $\cZ(\ZZ_3)$. 
This is precisely the $(E_6)_1 \times (A_2)_1$ theory. 

Each of the simple objects in $\cZ(\H_3)$ acts as a representation for the $\H_3$ symmetry. In particular, the decomposition of $(\mathfrak{e}_8)_1$ into representations of $\cH_3$ involves $1, \pi_1,$ and $\pi_2$  in the 
untwisted sector, and the action of $\alpha$, $\rho$ on these representations is the following,
\begin{itemize}
\item 1: one-dimensional representation with eigenvalues $\alpha =1$ and $\rho = d_\rho$. 
\item $\pi_1$: one-dimensional representation with eigenvalues $\alpha = 1 $ and $\rho= -d_\rho^{-1}$. 
\item $\pi_2$: two-dimensional representation with $\alpha =\left( \begin{smallmatrix} \omega & 0 \\ 0 & \omega^2\end{smallmatrix}\right)$ and $\rho$ acting as a permutation $\rho = \left( \begin{smallmatrix} 0 & 1 \\ 1 & 0 \end{smallmatrix}\right)$.
\end{itemize}
These can be obtained as special cases of the tube algebra described below.

As can be seen from e.g. the results in (16) of the main text, 
the $\alpha$ and $\alpha^2$ twisted sectors contain $\sigma_i$ for $i=1,2,3$, each in a one-dimensional representation, while the $\rho$, $\alpha \rho$, and $\alpha^2 \rho$ sectors contain all representations except the identity, each one again being one-dimensional. 
The action of the Haagerup symmetry on these twisted sector states is in general given by the following lasso action,
\bea
\label{eq:lassodiagram}
\begin{tikzpicture}[scale=0.4,baseline=15]
\begin{scope}[ thick, every node/.style={sloped,allow upside down}]
\draw[->- = 0.7](0,0) -- (0,1);
\draw[->-=0.7] (0,1) -- (0,2);
\draw[->- = 0.8](0,2) -- (0,3);
\draw(0,1) to[out=0,in=0,distance=0.6in] node{\opmidarrow}  (0,-1);
\draw  (0,-1) to[out=180,in=180,distance=0.8in] node{\opmidarrow}   (0,2);
\node[above] at (0,3) {\scriptsize $c$};
\node[left] at (-1.3,0) {\scriptsize $x$};
\node[right] at (0,1.4) {\scriptsize $b$};
\node[right] at (0.1,0.4) {\scriptsize$a$};
\node[below] at (0,0) {\scriptsize$\footnotesize\cO$};

\filldraw (0,0) circle (0.3ex);
\filldraw (0,1) circle (0.2ex);
\filldraw (0,2) circle (0.2ex);

\end{scope}
\end{tikzpicture}
\hspace{0.2 in}=\hspace{0.3 in}
\begin{tikzpicture}[scale=0.4,baseline=15]
\begin{scope}[ thick, every node/.style={sloped,allow upside down}]
\draw[->- = 0.7](0,0) -- (0,3);
\filldraw (0,0) circle (0.3ex);
\node[above] at  (0,3) {\scriptsize$c$};
\node[below] at (0,0) {\scriptsize$\footnotesize\cO'$};
\end{scope}
\end{tikzpicture}\hspace{0.1 in},
\eea
which can be computed by solving the tube algebra,\footnote{Other discussions of the tube algebra can be found in e.g. \cite{Lin:2022dhv,Choi:2024tri}.} 
\bea
\label{eq:lassodiagram}
\begin{tikzpicture}[scale=0.4,baseline=15]
\begin{scope}[ thick, every node/.style={sloped,allow upside down}]
\draw[->- = 0.7](0,0) -- (0,1);
\draw[->-=0.7] (0,1) -- (0,2);
\draw[->- = 0.8](0,2) -- (0,3);
\draw(0,1) to[out=0,in=0,distance=0.6in] node{\opmidarrow}  (0,-1);
\draw  (0,-1) to[out=180,in=180,distance=0.8in] node{\opmidarrow}   (0,2);

\draw(0,3) to[out=0,in=0,distance=1.2in] node{\opmidarrow}  (0,-2);
\draw  (0,-2) to[out=180,in=180,distance=1.4in] node{\opmidarrow}   (0,4);

\draw[->- = 0.8](0,3) -- (0,4);
\draw[->- = 0.8](0,4) -- (0,5);

\node[above] at (0,5) {\scriptsize $e$};
\node[right] at (0,3.5) {\scriptsize $d$};
\node[right] at (0,2.4) {\scriptsize $c$};
\node[left] at (-1.3,0) {\scriptsize $x$};
\node[left] at (-2.5,0.5) {\scriptsize $y$};
\node[right] at (0,1.4) {\scriptsize $b$};
\node[right] at (0,0.4) {\scriptsize$a$};
\node[below] at (0,0) {\scriptsize$\footnotesize\cO$};

\filldraw (0,0) circle (0.3ex);
\filldraw (0,1) circle (0.2ex);
\filldraw (0,2) circle (0.2ex);

\end{scope}
\end{tikzpicture}
&=&\kappa_y \sum_{f,g}\sqrt{d_x d_y^3 d_c d_f \over d_b d_d d_g} (F_{c y \bar y}^c)_{d1} (F_{x d \bar y}^b)_{fc} 
\no\\
&\vphantom{.}& \hspace{-0.7 in} \times(F_{f \bar y y}^f)^{-1}_{1b}  (F_{xye}^f)_{gd} (F_{axy}^f)_{bg}
\label{eq:lassodiagram}
\begin{tikzpicture}[scale=0.4,baseline=15]
\begin{scope}[ thick, every node/.style={sloped,allow upside down}]
\draw[->- = 0.7](0,0) -- (0,1);
\draw[->-=0.7] (0,1) -- (0,2);
\draw[->- = 0.8](0,2) -- (0,3);
\draw(0,1) to[out=0,in=0,distance=0.6in] node{\opmidarrow}  (0,-1);
\draw  (0,-1) to[out=180,in=180,distance=0.8in] node{\opmidarrow}   (0,2);
\node[above] at (0,3) {\scriptsize $e$};
\node[left] at (-1.3,0) {\scriptsize $g$};
\node[right] at (0,1.4) {\scriptsize $f$};
\node[right] at (0.1,0.4) {\scriptsize$a$};
\node[below] at (0,0) {\scriptsize$\footnotesize\cO$};

\filldraw (0,0) circle (0.3ex);
\filldraw (0,1) circle (0.2ex);
\filldraw (0,2) circle (0.2ex);

\end{scope}
\end{tikzpicture}~,
\eea
as follows from a straightforward sequence of $F$-moves. Note that $\kappa_y$ above is the Frobenius-Schur indicator of $y$, which is trivial for all elements of $\H_3$.
In practice, solving these equations is a complicated process and must be done with computer assistance. 
 Examples of the final results include
\bea
\label{eq:lassodiagram}
\begin{tikzpicture}[scale=0.4,baseline=15]
\begin{scope}[ thick, every node/.style={sloped,allow upside down}]
\draw[->- = 0.7](0,0) -- (0,1);
\draw[->-=0.7] (0,1) -- (0,2);
\draw[->- = 0.8](0,2) -- (0,3);
\draw(0,1) to[out=0,in=0,distance=0.6in] node{\opmidarrow}  (0,-1);
\draw  (0,-1) to[out=180,in=180,distance=0.8in] node{\opmidarrow}   (0,2);
\node[above] at (0,3) {\scriptsize $\rho$};
\node[left] at (-1.3,0) {\scriptsize$\rho$};
\node[right] at (0,1.4) {\scriptsize$\rho$};
\node[right] at (0.1,0.4) {\scriptsize$\rho$};
\node[below] at (0,0) {\scriptsize$\pi_1$};

\filldraw (0,0) circle (0.3ex);
\filldraw (0,1) circle (0.2ex);
\filldraw (0,2) circle (0.2ex);

\end{scope}
\end{tikzpicture}
\hspace{0.2 in} = -\sqrt{2 \over 61 + 17 \sqrt{13}} \hspace{0.3 in}
\begin{tikzpicture}[scale=0.4,baseline=15]
\begin{scope}[ thick, every node/.style={sloped,allow upside down}]
\draw[->- = 0.7](0,0) -- (0,3);
\filldraw (0,0) circle (0.3ex);
\node[above] at  (0,3) {\scriptsize$\rho$};
\node[below] at (0,0) {\scriptsize$\pi_1$};
\end{scope}
\end{tikzpicture}\hspace{0.1 in},
\eea
\bea
\label{eq:lassodiagram}
\begin{tikzpicture}[scale=0.4,baseline=15]
\begin{scope}[ thick, every node/.style={sloped,allow upside down}]
\draw[->- = 0.7](0,0) -- (0,1);
\draw[->-=0.7] (0,1) -- (0,2);
\draw[->- = 0.8](0,2) -- (0,3);
\draw(0,1) to[out=0,in=0,distance=0.6in] node{\opmidarrow}  (0,-1);
\draw  (0,-1) to[out=180,in=180,distance=0.8in] node{\opmidarrow}   (0,2);
\node[above] at (0,3) {\scriptsize $\rho$};
\node[left] at (-1.3,0) {\scriptsize$\rho$};
\node[right] at (0,1.4) {\scriptsize$\rho$};
\node[right] at (0.1,0.4) {\scriptsize$\rho$};
\node[below] at (0,0) {\scriptsize$\pi_2$};

\filldraw (0,0) circle (0.3ex);
\filldraw (0,1) circle (0.2ex);
\filldraw (0,2) circle (0.2ex);

\end{scope}
\end{tikzpicture}
\hspace{0.2 in} = {2\over 3} \sqrt{-2 + \sqrt{13}} \hspace{0.3 in}
\begin{tikzpicture}[scale=0.4,baseline=15]
\begin{scope}[ thick, every node/.style={sloped,allow upside down}]
\draw[->- = 0.7](0,0) -- (0,3);
\filldraw (0,0) circle (0.3ex);
\node[above] at  (0,3) {\scriptsize$\rho$};
\node[below] at (0,0) {\scriptsize$\pi_2$};
\end{scope}
\end{tikzpicture}\hspace{0.1 in},
\eea
and 
\bea
\label{eq:lassodiagram}
\begin{tikzpicture}[scale=0.4,baseline=15]
\begin{scope}[ thick, every node/.style={sloped,allow upside down}]
\draw[->- = 0.7](0,0) -- (0,1);
\draw[->-=0.7] (0,1) -- (0,2);
\draw[->- = 0.8](0,2) -- (0,3);
\draw(0,1) to[out=0,in=0,distance=0.6in] node{\opmidarrow}  (0,-1);
\draw  (0,-1) to[out=180,in=180,distance=0.8in] node{\opmidarrow}   (0,2);
\node[above] at (0,3) {\scriptsize $\rho$};
\node[left] at (-1.3,0) {\scriptsize$\rho$};
\node[right] at (0,1.4) {\scriptsize$\rho$};
\node[right] at (0.1,0.4) {\scriptsize$\rho$};
\node[below] at (0,0) {\scriptsize$\sigma_2$};

\filldraw (0,0) circle (0.3ex);
\filldraw (0,1) circle (0.2ex);
\filldraw (0,2) circle (0.2ex);

\end{scope}
\end{tikzpicture}
\hspace{0.2 in} = {1\over 3}\sqrt{17 + 5 \sqrt{13} \over 2}\, e^{- {2 \pi i \over 3}} \hspace{0.1 in}
\begin{tikzpicture}[scale=0.4,baseline=15]
\begin{scope}[ thick, every node/.style={sloped,allow upside down}]
\draw[->- = 0.7](0,0) -- (0,3);
\filldraw (0,0) circle (0.3ex);
\node[above] at  (0,3) {\scriptsize$\rho$};
\node[below] at (0,0) {\scriptsize$\sigma_2$};
\end{scope}
\end{tikzpicture}~.\hspace{0.1 in}
\eea
Data for all lassos can be made available upon request.

\section{Lifting Verlinde lines to non-diagonal modular invariants}
\label{app:lifts}

In the main text, we argued for the existence of symmetries in $(E_8)_1$ which do not commute with the chiral algebra, obtained by lifting Verlinde lines from theories related to $(E_8)_1$ via conformal embedding.  Here we will explain this idea in more detail. As a warm-up, we first discuss the more familiar case of the conformal embedding $(A_1)_4 \subset (A_2)_1$, and show that lifting the Verlinde lines of $(A_1)_4$ gives rise to two Tambara-Yamagami extensions of the $\ZZ_3$ center of $ (A_2)_1$. We then show how the Verlinde lines of the $\cZ(H_3)$ theory can be lifted to $(E_8)_1$ to give $\cH_3 \times \cH_3^\mathrm{op}$, and to $(E_6)_1 \times (A_2)_1$ to give a double near-group extension of $\ZZ_3^2$. 

Before describing these results, let us mention here two small checks that may be applied to each of our examples, 
\begin{enumerate}

\item Upon gauging an algebra $A \subset \cC$ of quantum dimension $d_A$ to obtain a fusion category $\widehat \cC$, there must exist a dual algebra object $\widehat A \subset \widehat \cC$ with $d_{\widehat A} = d_{ A}$. This is easily seen by writing $A = \cI \cI^*$ for $\cI$ an interface between $\cC$ and $\widehat \cC$, in which case $\widehat A = \cI^* \cI$; see e.g. \cite{Diatlyk:2023fwf}. 

\item The total quantum dimension of the categorical symmetry commuting with a fixed chiral algebra must be preserved upon discrete gauging. This follows from the fact that they share the same SymTFT.  Note that it is important that we consider the \textit{same} chiral algebra before and after gauging. For example, as we will review below, the $SU(2)_4$ and $SU(3)_1$ theories are related by gauging, but if we consider only the symmetries  which commute with the maximally extended chiral algebras $SU(2)$ and $SU(3)$ (i.e. the Verlinde lines), we find that they have total quantum dimension $12$ and $3$, respectively. 
On the other hand, if we allow for lines in the $SU(3)_1$ theory which do not commute with the full $SU(3)$, but rather only an $SU(2) \subset SU(3)$, then we find that these again have total quantum dimension $12$. 

\end{enumerate}
All of the results below are consistent with these facts.\footnote{The method we describe here can be thought of as a rephrasing of the method of $\alpha$-induction \cite{Longo:1994xe,xu1998new,
bockenhauer1998modular,bockenhauer1999modular,bockenhauer1999modular3,bockenhauer2000chiral,ostrik2003module,Komargodski:2020mxz}. }
\subsection{Warm-up: from $(A_1)_4$ to $(A_2)_1$ }

The diagonal $SU(2)_4$ theory has five primaries, with conformal dimensions $h = 0, {1/ 8}, {1/ 3},{5 / 8},  1$, and correspondingly five Verlinde lines $1, \cN, W, \eta \cN, \eta$ with the following fusion ring, 
\bea
&\vphantom{.}& \eta^2 = 1~, \hspace{0.1 in} \cN^2 = 1+ W~, \hspace{0.1 in} W^2 = 1 + \eta + W~, 
\no\\
&\vphantom{.}& \hspace{0.3 in}\eta W = W~, \hspace{0.1in} W \cN = \cN + \eta \cN~.
\eea
The action of these operators on the primaries can be computed by using the $S$-matrix 
\bea
S = {1\over 2 \sqrt{3}}\left(\begin{smallmatrix}  1 & \sqrt{3} & 2 & \sqrt{3} & 1 \\ \sqrt{3} & \sqrt{3} & 0 & - \sqrt{3} & -\sqrt{3} \\ 2 & 0 & -2 & 0 & 2 \\ \sqrt{3} & - \sqrt{3} & 0 & \sqrt{3} & -\sqrt{3} \\ 1 & -\sqrt{3} & 2 & -\sqrt{3} & 1\end{smallmatrix}  \right) 
\eea
and the following formula, 
\bea
\cL_i |j\rangle = {S_{ij} \over S_{0j}} |j \rangle~,
\eea
where $\cL_i$ is the $i$-th Verlinde line and $|j \rangle$ is the state corresponding to the $j$-th primary,
giving the following results,
\begin{equation}
\label{eq:table}
\begin{tabular}{c|ccccc}
& $0$ & ${1\over 8}$ & ${1\over 3}$ & ${5\over 8}$ & $1$
\\ \hline
$\eta$ & $1$ & $-1$ & $1$ & $-1$ & $1$ 
\\
$\cN$ &  $\sqrt{3}$ & $1$ & $0$ & $-1$ & $-\sqrt{3}$
\\
$W$ & $2$ & $0$ & $-1$ & $0$ & $2$
\end{tabular}~.
\end{equation}

Upon gauging the $\ZZ_2$ symmetry generated by $\eta$, we obtain the $SU(3)_1$ theory, with partition function given in terms of $SU(2)_4$ characters as 
\bea
Z_{SU(3)_1} = |\chi^{SU(2)_4}_0 + \chi^{SU(2)_4}_1|^2 + 2 |\chi^{SU(2)_4}_{1/ 3}|^2~.
\eea
As mentioned in the main text, we may now lift the non-invertible symmetries of $SU(2)_4$ to $SU(3)_1$. 
This lifting is carried out as follows. We first use the conformal embedding to define \textit{chiral} symmetries. 
In particular, let us write our theory in terms of the gluing matrix $M$ as
\begin{equation}
    Z_{SU(3)_1} = \chi M \bar{\chi}^T~,\hspace{0.2 in} M =  \left(\begin{smallmatrix} 1 & 0 & 0 & 0 & 1 \\ 0 & 0 & 0 & 0 & 0 \\ 0 & 0 & 2 & 0 & 0 \\ 0 & 0 & 0 & 0 & 0\\ 1 & 0 & 0 & 0 & 1\end{smallmatrix} \right);
\end{equation}
it is easy to check that this $M$ is modular invariant, i.e.
\begin{equation}
\label{eq:modularinvariantreq}
    M S= S M, \qquad M T = T M~,
\end{equation}
as it must always be.
Then denoting by $D_{\cL}$ for $\cL \in \cC$ the diagonal matrices whose entries are given by the rows in (\ref{eq:table}), i.e. $(D_{\cL_i})_{jj} = S_{i j} / S_{0 j}$,  we define the action of the chiral symmetries via 
\bea
Z_{SU(3)_1} |^{\cL_L} &=&  \chi^T D_{\cL} M  \overline \chi~,
\no
\eea
and similarly for the anti-chiral symmetries. These are guaranteed to give non-negative integer coefficients upon $S$-transformation, since by modular invariance of $M$ and symmetricity of $S$ we have 
\bea
  \chi^T D_{\cL} M  \overline\chi \,\,\rightarrow\,\,   \chi^T S D_\cL M S^* \overline  \chi = \chi^T  S D_\cL S^{*}  M \overline \chi ~,
 \no\\
\eea
and using the definition of  $(D_{\cL})_{jj}$  we have $(S D_\cL S^*)_{ij} = N_{\cL i}^j$ by the Verlinde formula. 
In other words, 
\bea
Z_{SU(3)_1} |_{\cL_L} &=&  \chi^T N_{\cL} M \, \overline\chi~.
\eea
Thus the chiral symmetries give rise to well-defined twisted Hilbert spaces. In some sense this is rather obvious---it is simply the statement that the symmetry actions continue to be defined for non-trivial modular invariants.

From this simple example, we see the following phenomenon: for non-trivial modular invariants, some of the symmetries can be absorbed into the gluing matrix $M$, the interpretation being that they are gauged (in the bulk language, it means they are condensed on the surface defect corresponding to $M$). 
Indeed, it is straightforward to see that $D_{\eta}  M = M$ for the modular invariant above. More generally, the matrices $D_\cL$ are organized into equivalence classes $[\cL]$ with equivalence relation $\cL_1 \sim \cL_2$ if and only if $D_{\cL_1} M =  D_{\cL_2} M$. Using this, we may now obtain the reduced fusion algebra as follows. First note that on the equivalence classes just defined, the fusion rules become 
\bea
&\vphantom{.}& [W]^2 = 2 [1]+ [W]~, \hspace{0.1 in} [\cN]^2 = [1] + [W]~,
\no\\
&\vphantom{.}&  \hspace{0.4 in} [\cN] [W] = [W] [\cN]=  2 [\cN]~.
\eea
The fact that 2 appears in front of $[1]$ in the first equation implies that $[W]$ is non-simple, and indeed since it is invariant under the $\eta$ gauging it should split, $[W] = [W_+] + [W_-]$.  Then the fusion rules become 
\bea
&\vphantom{.}& \hspace{0.3 in}  [W_\pm]^2 = [W_\mp]~, \hspace{0.1 in} [W_+][W_-] = [1]~,
\no\\
&\vphantom{.}& \hspace{-0.1 in} [\cN]^2 = [1]+[W_+]+[W_-]~, \hspace{0.1 in} [W_\pm] [\cN]= [\cN]~.\hspace{15pt}
\eea
Thus the chiral symmetry in the $SU(3)_1$ theory is simply $TY(\ZZ_3)$, with the duality defect descending from $\cN$ in $SU(2)_4$ and the $\ZZ_3$ symmetry descending from $W$. Noting that $\mathrm{diag}(2,0,-1,0,2) = \mathrm{diag}(1,0,e^{2 \pi i \over 3},0,1) + \mathrm{diag}(1,0,e^{-{2 \pi i \over 3}},0,1)$, it is easy to see that the $\ZZ_3$ appearing here is none other than the center of $SU(3)_1$. 
The fact that there is a Tambara-Yamagami extension of the center has been noticed previously in \cite{Bockenhauer:1998in,Damia:2024xju}.

Note that since $D_{W_\pm} M = M D_{W_\pm} $, there is only a single $\ZZ_3$ arising from the lift---the chiral and anti-chiral definitions give the same result. 
On the other hand, $D_{\cN} M \neq M D_{\cN} $  so we may really define two symmetries $[\cN_{L,R}]$, acting separately on chiral and anti-chiral modes.  
For example, the respective twisted sector partition functions are 
\bea
Z_{SU(3)_1} |_{\cN_R} &=& ( \chi_0^{SU(2)_4} +  \chi_1^{SU(2)_4} + 2  \chi_{1/3}^{SU(2)_4})
\no\\
&\vphantom{.}& \hspace{0.7 in} \times (\overline\chi_{1/8}^{SU(2)_4}+\overline\chi_{5/8}^{SU(2)_4})~, 
\no\\
Z_{SU(3)_1} |_{\cN_L} &=& ( \chi_{1/8}^{SU(2)_4}+ \chi_{5/8}^{SU(2)_4})
\\
&\vphantom{.}& \hspace{0.1 in} \times( \overline\chi_0^{SU(2)_4} +  \overline\chi_1^{SU(2)_4} + 2 \overline \chi_{1/3}^{SU(2)_4}) ~. 
\no
\eea
It remains only to study their product $[\cN_L][\cN_R]$. It is easy to see that this object is not simple---indeed, we have 
\bea
([\cN_L ][\cN_R])^2 = 3 \left( [1]+[W_+]+[W_-]\right)~, 
\eea
and hence the product should be expanded as the sum of three simple lines. From the quantum dimension, it is clear that these lines must be invertible, and hence we expect to have $[\cN_L][\cN_R]=[\mathsf{C}]( [1]+[W_+]+[W_-])$ , where $[\mathsf{C}]$ is a $\ZZ_2$ element. What this means is that we should really think of $[\cN_R]$ as a product $[\mathsf{C}] [\cN_L]$. The symmetry $[\mathsf{C}]$ does not descend from the lines in the original theory in the same way as e.g. $[\cN_{L,R}]$ did---it is the quantum $\ZZ_2$ symmetry appearing after gauging $\eta$. It is this $\ZZ_2$ symmetry that we must gauge in order to return to the $SU(2)_4$ theory. 

In total then, the objects obtained from lifting the Verlinde lines in $SU(2)_4$ are as follows,  
\bea
[1],~[ \mathsf{C}],~ [W_+],~[W_-] ,~ [\mathsf{C}W_+],~ [\mathsf{C}W_-],~ [\cN_L],~ [\mathsf{C}\cN_L]~. 
\no\\
\eea
The total quantum dimension here is $\cD^2 = 6 + 2 (\sqrt{3})^2 =12$,  which is the same as the original total quantum dimension of the MTC of $SU(2)_4$. 

\subsection{From $\cZ(\H_3)$ to $(E_8)_1$}
\label{app:E8lifts}

We now repeat this analysis to show how the Verlinde lines of the $\cZ(\H_3)$ theory can be lifted to the $(E_8)_1$ theory to give $\H_3 \times \H_3^\mathrm{op}$. The modular gluing matrix in this case is\footnote{There are a total of 28 non-negative integer solutions $M$ to the modular invariance equations (\ref{eq:modularinvariantreq}) for $\cZ(\H_3)$, see e.g. \cite{evans2006modular}. }
\bea
M = \left( \begin{smallmatrix} 
1 & 1 & 2 & 0 &  \dots & 0
\\
1& 1 & 2 & 0 &   \dots & 0
\\
2 & 2 & 4 &  0 &  \dots & 0
\\
0 & 0 & 0 &0 &   \dots & 0
\\
\vdots &  &  &  &  \ddots & \vdots
\\
0 & 0 & 0 & 0 &   \dots & 0
\end{smallmatrix} \right) ~,
\eea
and we define the action of the chiral symmetry in terms of the $S$-matrix elements via $(D_{\cL_i})_{jj} = S_{i j} / S_{0j}$ for $\cL_i \in \cZ(\H_3)$. 

As before, the first step is to determine the equivalence classes with respect to $M$. It is easy to see that $ D_{\sigma_i} M=  D_{\sigma_j} M$ for any $i,j = 1,2, 3$, and similarly that $ D_{\mu_i} M=  D_{\mu_j} M$ for any $i,j = 1, \dots, 6$. There are precisely two additional linear relations, namely
\bea
 D_{\pi_1}M &=& (D_1 + D_{\mu_i})M~, 
\no\\
 D_{\pi_2} M&=& (D_1 + D_{\pi_1})M~. 
\eea
We thus have three distinct equivalence classes, which we denote by $[1], [\sigma],$ and $[\mu]$. 
In particular, the would-be equivalence classes $[\pi_{1,2}]$ are expressible in terms of these as
\bea
[\pi_1 ] = [1 ] + [\mu]~, \hspace{0.2 in} [\pi_2] = 2[1]+[\mu]~. 
\eea

The next step is to express the fusion rules in terms of these equivalence classes. The original fusion rules for the $\cZ(\H_3)$ theory follow from the Verlinde formula, applied to the $S$-matrix in (\ref{eq:ZH3Smatrix}). Expressing these fusions in terms of the equivalence classes above, we obtain the following, 
\bea
\label{eq:H3fusionrulesfromZH3}
\, [\sigma]^2 &=& 5 [1] + 4 [\sigma] + 9 [\mu]~, 
\no\\
\, [\mu]^2 &=& 3 [1] + 3 [\sigma] + 6 [\mu]~, 
\no\\
\, [\sigma] [\mu] &=& 3 [1] + 3 [\sigma] + 8 [\mu]~.
\eea
In particular, one finds that
\bea
\label{eq:sigmaminusmu}
([\sigma] - [\mu])^2 = 2 [1] + ([\sigma] - [\mu])~.
\eea
The presence of the factor of $2[1]$ on the right-hand side implies that $[\sigma] - [\mu]$ is non-simple. Hence we split it as $[\sigma] - [\mu] = [\alpha]+[\alpha^2]$, for which the fusion rule in (\ref{eq:sigmaminusmu}) reduces to the usual one for a $\ZZ_3$ symmetry generated by $ [\alpha]$. 
Plugging this into the second equation of (\ref{eq:H3fusionrulesfromZH3}), we obtain 
\bea
[\mu]^2 = 3 [1] + 3[\alpha] + 3[\alpha^2] + 9 [\mu]~. 
\eea
The factor of $3[1]$ on the right-hand side implies that $[\mu]$ itself is also non-simple, and should be split as $[\mu] = [\rho_0] + [\rho_1]+[\rho_2]$. Plugging this into the above equations then gives fusion rules consistent with those of $\H_3$. 

Thus we see that the chiral symmetry induced by lifting the $\cZ(\H_3)$ Verlinde lines to $(E_8)_1$ is precisely $\H_3$. 
Clearly, the same thing can be done for the anti-chiral symmetry, which is easily checked to give a distinct copy of $\H_3^\mathrm{op}$. 
We conclude that the total symmetry obtained from lifting the $\cZ(\H_3)$ Verlinde lines is $\H_3 \times \H_3^\mathrm{op}$, acting on the chiral and anti-chiral $(\mathfrak{e}_8)_1$ halves.

\subsection{From $\cZ(\H_3)$ to $(E_6)_1 \times (A_2)_1$}
\label{app:E6A2lifts}

We next describe the lift of the Verlinde lines of the $\cZ(\H_3)$ theory to $(E_6)_1 \times (A_2)_1$. 
In this case the modular gluing matrix is 
\bea
M = \left( \begin{smallmatrix} 1 & 1 & 0 & 0 & 0 & 0 &  0 &\dots & 0
\\
1 & 1 & 0 & 0 & 0 & 0 & 0 &\dots & 0
\\
0 & 0 & 2 & 0 & 0 & 0 &  0 &\dots & 0
\\
0 & 0 & 0 & 2 & 0 & 0 &  0 &\dots & 0
\\
0 & 0 & 0 & 0 & 2 & 0&  0 & \dots & 0
\\
0 & 0 & 0 & 0 & 0 & 2 & 0 & \dots & 0
\\
0 & 0 & 0 & 0 & 0 & 0& 0 & \dots & 0
\\
\vdots &  &  &  &  &  &  & \ddots & \vdots
\\
0 & 0 & 0 & 0 & 0 & 0& 0 & \dots & 0
\end{smallmatrix} \right) ~. 
\eea
It is easy to check that $D_{\mu_i} M=  D_{\mu_j}M$ for any $i,j = 1, \dots, 6$ and there is precisely one additional linear relation, 
\bea
 D_{\pi_1} M&=& (D_1 + D_{\mu_i})M~. 
\eea
We thus have six equivalence classes $[1], [\pi_2], [\sigma_i],$ and $[\mu]$. For notational simplicity we denote $[\pi_2] = [\sigma_0]$ below. 
In terms of these equivalence classes, the fusion rules are written as follows,
\begin{equation}
\begin{split}
\,[\mu]^2 &= [1] + \sum_{i=0}^3 [\sigma_i] + 5 [\mu]~, \hspace{0.1 in} [\sigma_i][\mu] = [1] +  \sum_{i=0}^3 [\sigma_i] + 7 [\mu]~, 
\\
\,&\vphantom{.}\hspace{-0.4 in} [\sigma_i][\sigma_j] = \left\{ \begin{matrix} 
3 [1] + 2[\sigma_i] + \sum_{k \neq i} [\sigma_k] + 8 [\mu] & i=j
\\
 [1] + [\sigma_i] +[\sigma_j] +2 \sum_{k \neq i,j} [\sigma_k] + 7 [\mu] & i\neq j
\end{matrix} \right. ~.
\\
\end{split}
\end{equation}
From these we  see that 
\bea
([\sigma_i] - [\mu])^2 = 2 [1] + ([\sigma_i] - [\mu])~
\eea
for all $i=0,\dots, 3$, so each of the differences is non-simple, and should be decomposed into the products of generators of two $\ZZ_3$ symmetries, i.e. $[\sigma_i] - [\mu] = [g_i] + [g_i^2]$ where each of $g_i$ can be written as $\alpha_1^n \alpha_2^m$ for some $n,m$ and $\alpha_{1,2}$ the generators of the two $\ZZ_3$ symmetries. This then gives the fusions 
\bea
\label{eq:neargroupfusions}
[\mu]^2 
= 9 [\mu]+ \sum_{g \in \ZZ_3^2} [g]~, 
\hspace{0.1 in} [\mu][g]&=&[g][\mu]=[\mu]~, \hspace{15pt}
\eea
from which we conclude that $[\mu]$ is a simple line. The chiral symmetry induced in $(E_6)_1 \times (A_2)_1$ from the Verlinde lines in $\cZ(\H_3)$ is thus a near-group extension of $\ZZ_3^2$ by the non-invertible operator $[\mu]$. These near-group fusion rules admit a unique categorical completion,  denoted by $\ZZ_3^2 + 9$ in \cite{Evans:2012ta}. 

It is straightforward to check that $M D_{g_i} = D_{g_i} M$, so the anti-chiral symmetry does not give new $\ZZ_3$ factors. 
On the other hand, $M D_{\mu} \neq D_{\mu} M$, and hence we have separate near-group extensions on the left and right. We now consider the product $[\mu_L] [\mu_R]$. As in the case of $SU(2)_4 \subset SU(3)_1$, by computing the square it is easy to see that this is not simple, and that it must be split into the sum of 9 simple lines. Based on the quantum dimension, we then expect to obtain 
\bea
[\mu_L] [\mu_R] = \left( \sum_{g \in \ZZ_3^2} [g]\right) [\lambda]~,
\eea
where $[\lambda]$ is a line of dimension $d_\rho^2$. Computing $([\mu_L] [\mu_R])^2$ using (\ref{eq:neargroupfusions}) and the above expression then gives e.g.
\bea
[\lambda]^2 = [1]+9[\lambda] + [\mu_L] + [\mu_R]~.
\eea
It is this object $[\lambda]$ that we must gauge to go back to the $\cZ(\H_3)$ theory. 

To summarize, lifting the Verlinde lines in $\cZ(\H_3)$ to $(E_6)_1 \times (A_2)_1$ gives the following simple objects, 
\bea
[g]~,~~ [g \lambda]~,~~ [\mu_L]~,~~ [\mu_R]~,\hspace{0.2 in} g \in \ZZ_3^2~,
\eea
with total quantum dimension $9 + 9 (d_\rho^2)^2 + 2(3 d_\rho)^2 = {117\over 2}(11 + 3\sqrt{13}) $, matching with the total quantum dimension of $\cZ(\H_3)$.\footnote{Alternatively, starting from $(E_8)_1$ we can gauge $\ZZ_{3\mathrm{B}}$ to go to $(E_6)_1 \times (A_2)_1$, and hence we expect that the above is equivalent to $\H_3 \times \H_3^\mathrm{op}/\ZZ_{3\mathrm{B}}$. A similar relationship between Haagerup and $\ZZ_3^2 + 9$ seems to have been previously noted by Izumi, see \cite[Section 12]{izumi2015cuntz} and \cite[slides 19,20]{Izumislides}.}

\subsection{Additional examples: $\mathrm{Fib} \times \mathrm{Fib}^\mathrm{op}$}

\label{app:Fib}

Let us close by briefly mentioning some additional examples. We begin with the conformal embedding $(G_2)_1 \times (F_4)_1 \subset (E_8)_1$, with the corresponding relation between characters,
\bea
\label{eq:E8G2F4rel}
\chi_0^{E_8} = \chi_0^{G_2} \chi_0^{F_4} + \chi_{2/5}^{G_2} \chi_{3/5}^{F_4}~. 
\eea
The gluing and $S$-matrices are given by 
\bea
M = \left(\begin{smallmatrix} 1 & 0 & 0 & 1 \\ 0 & 0 & 0 & 0 \\ 0 & 0 & 0 & 0 \\ 1 & 0 & 0 & 1 \end{smallmatrix}  \right)~, \hspace{0.1 in} S ={1\over 1 + \phi^2}\left(\begin{smallmatrix} 1& \phi & \phi& \phi^2 \\ \phi & -1 & \phi^2 & - \phi \\ \phi & \phi^2 & -1 & - \phi \\ \phi^2 & - \phi & - \phi &. 1 \end{smallmatrix}  \right) ~ ~~~~~~
\eea
where $\phi = \half(1+ \sqrt{5})$ is the golden ratio. Starting with the four Verlinde lines $1, W_1, W_2, W_1 W_2$ with standard Fibonacci fusion rules $W_i^2 = 1+ W_i$, we now compute the equivalence classes with respect to $M$. It is easy to check that $D_{W_1}M  =  D_{W_2}M $ and 
\bea
\label{eq:FibFibopexample}
 D_{W_1 W_2} M  = D_{1} M +  D_{W_1}M~, 
\eea
and hence we obtain just two distinct equivalence classes $[1], [W_L]$ which generate a Fibonacci symmetry. 
Similar comments hold for the anti-chiral symmetry, with the same fusion rules. 
In total then, we find that the Verlinde lines in  $(G_2)_1 \times (F_4)_1$ lift to a $\mathrm{Fib}\times \mathrm{Fib}^{\mathrm{op}}$ category in $(E_8)_1$.\footnote{A similar observation was made in \cite{Rayhaun:2023pgc,Moller:2024xtt}. }

Note that the presence of the identity on the right-hand of (\ref{eq:FibFibopexample}) suggests that gauging $W_1 W_2$ should map us from    $(G_2)_1 \times (F_4)_1$ to $(E_8)_1$;  this is also implicit from the character relation (\ref{eq:E8G2F4rel}). To see this more explicitly, we may write $W:= W_1 W_2$ and consider gauging the binary algebra algebra object $\cA = 1+W$. The  (co)multiplication  may be chosen in such a way that the gauged torus partition function is simply, 
\bea
\label{eq:G2F4gauged}
Z_\mathrm{gauged} &=& {1 \over 1 + \phi^2} \left(Z_{G_2 \times F_4} + Z_{G_2 \times F_4}|_W+ Z_{G_2 \times F_4}|^W \right.
\no\\
&\vphantom{.}& \hspace{0.1 in}\left.  + Z_{G_2 \times F_4}|^W_{W,1} + Z_{G_2 \times F_4}|^W_{W,W} \right) ~.
\eea
The only term which is non-trivial to compute is the final one, since it cannot be expressed in terms of ($S$-transforms of) the others by a series of $F$-moves. However, it can be computed by explicitly solving the tube algebra, with the final result,
\begin{equation}
\begin{split}
Z_{G_2 \times F_4}|^W_{W,W} &= \phi\, \chi_0^{G_2} \chi_0^{F_4} \overline \chi_{2/5}^{G_2} \overline \chi_{3/5}^{F_4} + \phi \, e^{2 \pi i {2\over5}} \chi_{2/5}^{G_2} \chi_0^{F_4} \overline \chi_{0}^{G_2} \overline \chi_{3/5}^{F_4} 
\\
&\vphantom{.}  - \phi^{-1} \, e^{2 \pi i {1\over5}} \chi_{2/5}^{G_2} \chi_0^{F_4} \overline \chi_{2/5}^{G_2} \overline \chi_{3/5}^{F_4} 
\\
&\vphantom{.}- \phi^{-1} \, e^{-2 \pi i {1\over5}} \chi_{0}^{G_2} \chi_{3/5}^{F_4} \overline \chi_{2/5}^{G_2} \overline \chi_{3/5}^{F_4} 
\\
&\vphantom{.} + \mathrm{c.c} + \phi^{-3}|\chi_{2/5}^{G_2} \chi_{3/5}^{F_4}|^2~.
\end{split}
\end{equation}
Plugging this into (\ref{eq:G2F4gauged}) and using (\ref{eq:E8G2F4rel}) confirms that $Z_\mathrm{gauged}  = |\chi^{E_8}_0|^2$. 

Conversely, starting in the $(E_8)_1$ theory, we may gauge the diagonal $[W] := [W_L] [ W_R]$ to go back to the $(G_2)_1 \times (F_4)_1$  WZW model; the contribution $Z_{E_8}|_{[ W], [ W]}^{[W]}$ is again computed by solving the tube algebra explicitly.

Let us  note in passing that $(E_6)_1$ can also be shown to have a  $\mathrm{Fib} \times \mathrm{Fib}^\mathrm{op}$ symmetry. One way to do so is to use fact that $(E_6)_1$ can be related by non-invertible gauging to $(F_4)_1 \times (\mathrm{3\text{-}State\,\,Potts})$; at the level of characters, 
\bea
\chi_0^{E_6} &=& \chi_0^{F_4} \chi_0^{3 \mathrm{SP}}+ \chi_{3/5}^{F_4} \chi_{2/5}^{3 \mathrm{SP}}~,
\no\\
\chi_{2/3}^{E_6} &=& \chi_0^{F_4} \chi_{2/3}^{3 \mathrm{SP}}+ \chi_{3/5}^{F_4} \chi_{1/15}^{3 \mathrm{SP}}~. 
\eea
By repeating the usual analysis, one finds that the chiral symmetry induced from lifting the Verlinde lines in $(F_4)_1 \times (\mathrm{3\text{-}State\,\,Potts})$ is simply the $\mathrm{3\text{-}State\,\,Potts}$ fusion category itself, i.e. a set of six lines $1$, $\alpha$, $\alpha^2$, $W$, $\alpha W$, $\alpha^2 W$, with $\alpha$ generating a $\ZZ_3$ symmetry (which is the center of $E_6$) and $W$ having Fibonacci fusion rules. Another copy of $W$ appears from the anti-chiral symmetry, so that in total we obtain the anticipated $\mathrm{Fib} \times \mathrm{Fib}^\mathrm{op}$.

\subsection{Additional examples: from $(G_2)_3$ to $(E_6)_1$ }

Finally, consider the conformal embedding of $(G_2)_3 \subset (E_6)_1$. The $(G_2)_3$ theory has six primaries of dimensions $0, 2/7, 2/3, 4/7,8/7,1$. 
In this order, the modular $S$-matrix is given by \cite{Coquereaux:2010we}
\bea
S={1\over \sqrt{21}}\left(\begin{smallmatrix} 
{\sqrt{5 - \sqrt{21} \over 2}} & \sqrt{3} & \sqrt{7} & \sqrt{3}& \sqrt{3}& {\sqrt{5 + \sqrt{21} \over 2}}   
\\
\sqrt{3} & \sqrt{3}  u_3 & 0 &  \sqrt{3}  u_2 &  \sqrt{3}  u_1 & -\sqrt{3}
\\
\sqrt{7} &0 & -\sqrt{7} & 0 & 0 & \sqrt{7}
\\
\sqrt{3} & \sqrt{3}  u_2 & 0 &  \sqrt{3}  u_1 &  \sqrt{3}  u_3 & -\sqrt{3}
\\
\sqrt{3} & \sqrt{3}  u_1 & 0 &  \sqrt{3}  u_3 &  \sqrt{3}  u_2 & -\sqrt{3}
\\
{\sqrt{5 +\sqrt{21} \over 2}} &- \sqrt{3} & \sqrt{7} &- \sqrt{3}&- \sqrt{3}& {\sqrt{5 - \sqrt{21} \over 2}}  
\end{smallmatrix} \right)
\no\\
\eea
where $u_n := -2 \mathrm{cos}{2 n \pi \over 7}$, and the modular gluing matrix giving the $(E_6)_1$ theory takes the form
\bea
M = \left( \begin{smallmatrix} 
1& 0 & 0 & 0 & 0 & 1 \\
0& 0 & 0 & 0 & 0 & 0\\
0& 0 & 2 & 0 & 0 & 0\\
0& 0 & 0 & 0 & 0 & 0\\
0& 0 & 0 & 0 & 0 & 0\\
1& 0 & 0 & 0 & 0 & 1\\
 \end{smallmatrix} \right) ~. 
\eea
It is easy to verify that there are three distinct equivalence classes, which we denote by $[1], [X],$ and $[Y]$, with  fusion rules given by 
\bea
\label{eq:G23example}
\,[X]^2 &=& [1]+2 [X]+[Y]~,
\no\\
\,[Y]^2 &=& 3 [1] + 5 [X] + 2 [Y]~, 
\no\\
\,[X][Y]&=& [1] + 4 [X] + [Y]~. 
\eea 
From these we see that 
\bea
([Y]-[X])^2 = 2 [1] + ([Y]-[X])~, 
\eea
and hence we conclude that $[Y]-[X] = [\alpha] + [\alpha^2]$ with $ [\alpha]$ generating a $\ZZ_3$ symmetry. 
Plugging this back into (\ref{eq:G23example}) gives the fusion rules 
\bea
\,[X]^2 = 3 [X] + \sum_{\alpha \in \ZZ_3}[\alpha]~. 
\eea
In other words, the chiral symmetry forms a near-group category $\ZZ_3+3$, also known as the Izumi-Xu 2221 subfactor (see e.g. \cite{han2010construction}).
The discussion here gives a significantly simpler understanding of the relationship between the 2221 subfactor and $(G_2)_3$ mentioned in \cite[Appendix A]{calegari2011cyclotomic}. 

Finally, note that the anti-chiral symmetry gives an analogous near-group extension. If we denote the two non-invertible symmetries by $[X_L]$, $[X_R]$, their product $[X_L][X_R]$ is non-simple, splitting into three simple lines of quantum dimension $1 + d_X$ for $d_X = {1\over 2}(3 + \sqrt{21})$, which we write as 
\bea
[X_L][X_R]= ([1]+ [\alpha] + [\alpha^2]) [Z]~. 
\eea
In summary, the simple lines we find are the following,
\bea
\,[g],~[g Z],~ [X_L],~[X_R]~,\hspace{0.1in} g \in \ZZ_3
\eea
with total quantum dimension $3+3(1+d_X)^2 + 2 d_X^2 = {21 \over 2}(5 + \sqrt{21})$, matching the total quantum dimension of the MTC of $(G_2)_3$.

\section{Modular bootstrap with non-invertible symmetries}
\label{app:modularbootstrap}

In this section, we summarize the modular bootstrap techniques used to obtain Figures 2 and 3 of the main text. These techniques were originally developed in \cite{Lin:2023uvm}. The basic idea is to apply the modular bootstrap not to the partition function of a single two-dimensional theory, but rather to the character vector, realized as a boundary state for a three-dimensional SymTFT. Denoting the elements of the SymTFT by $\mu \in \cZ(\cC)$,  the 3d solid torus partition function in the $\mathcal{V}_\mu$ sector is given by 
\begin{align}
Z_{\mu}^{3\mathrm{d}}(\tau, \bar{\tau}) =
\sum_{(h, \bar{h}) \in \mathcal{H}_{\mu}} 
n_{\mu; h, \bar{h}} \, \chi_h(\tau) \, \chi_{\bar{h}}(\bar{\tau})~,
\end{align}
where $n_{\mu; h, \bar{h}}$ is a non-negative integer and $\cH_\m$ is the defect Hilbert space associated to $\m$. The spins $s:= h - \bar h$ of states in $\cH_\m$ are constrained in terms of the $T$ matrix entries by the spin-selection rule \cite{Chang:2018iay}. 

For $c>1$, the Virasoro characters take the following form,
\begin{align}
\chi_h(\tau) = \frac{q^{(h - \frac{c-1}{24})}}{\eta(\tau)}, \qquad \text{for } h > 0~,
\end{align}
and
\begin{align}
\chi_0(\tau) = \frac{q^{- \frac{c-1}{24}}}{\eta(\tau)} (1 - q (\tau))~.
\end{align}

From the 3D perspective, the modular $S$ crossing equation can be written as,
\begin{align}
Z_{\mu}^{3\mathrm{d}}\left(-\frac{1}{\tau}, -\frac{1}{\bar{\tau}}\right)
=
\sum_{\nu \in \mathcal{Z}(\mathcal{C})}
S_{\mu\nu}
Z_{\nu}^{3\mathrm{d}}(\tau, \bar{\tau})~.
\end{align}
This can be equivalently rewritten as
\begin{align}
0 = \sum_{\nu \in \mathcal{Z}(\mathcal{C})} 
\sum_{(h, \bar{h}) \in \mathcal{H}_\nu} 
n_{\nu; h, \bar{h}}\, X_{\mu, \nu; h, \bar{h}}(\tau, \bar{\tau})~
\label{1}
\end{align}
for all anyons $\m$, 
where $X_{\mu, \nu; h, \bar{h}}(\tau, \bar{\tau})$ is defined as
\begin{align}
X_{\mu, \nu; h, \bar{h}}(\tau, \bar{\tau}) := 
I_{\mu \nu} \chi_h(-1/\tau)\chi_{\bar{h}}(-1/\bar{\tau}) 
- S_{\mu \nu} \chi_h(\tau)\chi_{\bar{h}}(\bar{\tau})~.
\end{align}

To test whether a putative spectrum satisfies the crossing equation \eqref{1}, we introduce a linear functional,
\begin{align}
\alpha_\mu := \sum_{m,n=0}^{\Lambda} \alpha_\mu^{m,n} 
\left. \partial_\tau^m \partial_{\bar{\tau}}^n \right|_{\tau = -\bar{\tau} = i}~, 
\qquad \alpha_\mu^{m,n} \in \mathbb{R}~,
\end{align}
with $\Lambda$ an appropriate upper bound.
Acting with this functional on \eqref{1} yields
\begin{align}
\hspace{-20pt}0 
&= 
\sum_{\mu,\nu \in \mathcal{Z}(\mathcal{C})} 
\sum_{(h, \bar{h}) \in \mathcal{H}_\nu} 
n_{\nu; h, \bar{h}}\, \alpha_\mu\left[ X_{\mu, \nu; h, \bar{h}} \right] \notag \\
&=
\sum_{\mu,\nu \in \mathcal{Z}(\mathcal{C})}
\sum_{(h, \bar{h}) \in \mathcal{H}_\nu} 
\sum_{m,n=0}^{\Lambda}
n_{\nu; h, \bar{h}} \
\alpha_\mu^{m,n} 
\left. \partial_\tau^m \partial_{\bar{\tau}}^n X_{\mu, \nu; h, \bar{h}} \right|_{\begin{smallmatrix}
    \tau = i\phantom{-}\\ \bar{\tau} = -i
\end{smallmatrix}}. \label{2}
\end{align}

A conformal field theory necessarily contains the vacuum state $(h, \bar{h}) = (0,0)$.
For the vacuum, we require that the vector $\alpha_\mu^{m,n}$ satisfies
\begin{align}
\sum_{\mu \in \mathcal{Z}(\mathcal{C})} 
\alpha_\mu\left[ X_{\mu, 1; 0, 0} \right]
>
0
, \quad
\text{for } (0, 0) \in \mathcal{H}_1 ~. \label{3}
\end{align}
Then if, given a putative operator spectrum, we are able to identify an $\alpha_\mu^{m,n}$ such that
\begin{align}
\sum_{\mu \in \mathcal{Z}(\mathcal{C})} 
\alpha_\mu\left[ X_{\mu, \nu; h, \bar{h}} \right]
\ge
0
, \quad
\text{for } \forall (h, \bar{h}) \in \mathcal{H}_\nu ~, \label{4}
\end{align}
then clearly \eqref{2} cannot be satisfied, and hence such a spectrum must be excluded. 

Note that when there exists a vector $v$ which is a simultaneous eigenvector of $S$ and $T$ with eigenvalue 1, and this vector has a zero component in the slot labelled by $\mu$, then this vector can be interpreted as a solution to the crossing equations with empty $\cV_\m$ sector, and hence there cannot be an upper bound on dimensions in the sector $\cV_\m$. In the case of Haagerup, the space of eigenvectors $v$ with eigenvalue 1 is spanned by 
\bea
v_1 &=& (1,1,2,0,0,0,0,0,0,0,0,0)~,
\no\\
v_2 &=& (1,1,0,2,0,0,0,0,0,0,0,0)~, 
\eea
as expected from modular invariance of (13) of the main text. From this, we see that upper bounds exist in $\cV_1$, $\cV_{\pi_1}$, and $\cV_{\pi_2} \cup \cV_{\sigma_1}$. Individually, $\cV_{\pi_2}$ or $\cV_{\sigma_1}$ alone would not admit an upper bound due to the presence of $v_2$ and $v_1$, respectively; however, the vector $(*,*,0,0,*,\dots,*)$ does not belong to this eigenspace, so $\cV_{\pi_2} \cup \cV_{\sigma_1}$ \textit{does} admit an upper bound.

Using the $S$ and $T$ matrices of $\cZ(\H_3)$ given in the Supplemental Materials, together with SDPB \cite{Simmons-Duffin:2015qma}, 
we are then able to obtain bounds on the lightest scalar operator in each of these three sectors.\footnote{In order to reduce \eqref{3} and \eqref{4} to a semi-definite programming problem, we list \eqref{4} for fixed spins $s$, and we truncate at $|s|~\le~\frac{s_{\mathrm{max}}}{2}$. To obtain our plots, we have taken $\Lambda = 21$, $s_{\mathrm{max}} = 16$, and \texttt{precision}$=320$.} The results are given in  Figures 2 and 3 for the sectors $\cV_{1}$ and $\cV_{\pi_2} \cup \cV_{\sigma_1}$, respectively. As we can see, the theory $(A_2)_1$ lies at a prominent kink in both plots. The theory $(E_6)_1$ sits at a kink in Figure 2, and is within the allowed region in Figure 3.

\section{More on symmetries in the chiral $(\mathfrak{e}_8)_1$ WZW model}
\label{app.chiral}

\begin{figure}[t]
\begin{tikzpicture}

     \node[right]  at (-2.3,0) {$\H_1$};
     \node[left] at (3.5,0) {$\H_2$};
      \node[]  at (0.5,2.1) {$\H_3$};

    \draw[-stealth,thick] (2.25,-0.05)--(-1.3,-0.05) node[midway,  below] {$1\oplus \tilde{\rho}$};
   \draw[-stealth,thick ] (-1.3,0.05)--(2.25,0.05) node[midway,  above] {$1\oplus\eta$};
        
  \draw[-stealth,thick] (-1.7+0.05,0.3)--(0.2+0.05-0.1,1.9) node[midway, sloped, below] {$1\oplus\mu\oplus\nu$};
  \draw[-stealth,thick] (0.2-0.1-0.1,1.9)--(-1.7-0.1,0.3) node[midway,sloped,  above] {$1\oplus\rho\oplus\alpha \rho$};
  
    \draw[-stealth,thick] (3-0.1-0.1,0.3)--(1-0.1-0.1,1.9) node[midway, sloped, below] {$1\oplus \tilde{\alpha}\oplus \tilde{\alpha}^2$};
  \draw[-stealth,thick] (1+0.05-0.1,1.9)--(3+0.05-0.1,0.3) node[midway, sloped, above] {$1\oplus\alpha \oplus \alpha^2$};

    \end{tikzpicture}
    \caption{The web of gaugings relating the $\H_1, \H_2, \H_3$ Haagerup symmetries. }
\label{fig:gaugings2}
\end{figure}
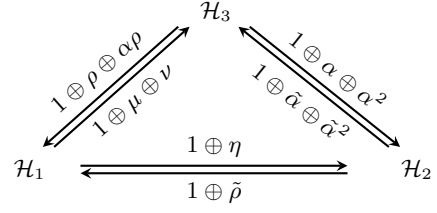

In this section, we provide some more details on the symmetries of the chiral $(\mathfrak{e}_8)_1$ WZW model. 
Let us denote the simple objects in $\H_1$ by $\{1, \nu, \eta, \mu\}$, those in $\H_2$ by $\{1, \tilde{\alpha}, \tilde{\alpha}^2, \tilde{\rho}, \tilde{\alpha} \tilde{\rho}, \tilde{\alpha}^2 \tilde{\rho}\}$, and those in $\H_3$ by $\{1,\alpha, \alpha^2, \rho, \alpha \rho, \alpha^2 \rho\}$.\footnote{In the main text, the lines in $\H_3$ were denoted  by $\{1,\widehat{\alpha}_L, \widehat{\alpha}_L^2, \widehat{\rho}_L, \widehat{\alpha}_L \widehat{\rho}_L, \widehat{\alpha}_L^2 \widehat{\rho}_L\}$, but in the current section we suppress the hat and subscript for simplicity. } The fusion rule of the objects in $\H_2$ are the same as those of $\H_3$, see (1) of the main text, while the fusion rules in $\H_1$ are given by \cite[Table 2]{Grossman_2012},
\bea
\eta^2 &=& 1 + \eta + \mu + \nu~, \hspace{0.35 in} \mu^2 = 1 + \nu~,
\no\\
 \nu^2 &=& 1 + 2 \eta + \mu + 2 \nu~, \hspace{0.2 in} \eta \mu = \eta + \nu~,
\no\\
 \eta \nu &=& \eta + \mu + 2 \nu~, \hspace{0.5 in} \mu \nu = \eta + \mu + \nu~.
\eea
 Each of the $\H_i$ fusion categories is related by gauging, as summarized in Figure \ref{fig:gaugings2}. As a consistency check, the two algebras between any given pair $(\H_i, \H_j)$ share the same quantum dimension. See \cite{Grossman_2012, Huang:2021ytb, Bottini:2025hri} for more details. 

Since the $\H_{i}$ are all related by gauging, they share the same SymTFT, whose category of lines is the Drinfeld center $\mathcal{Z}(\H_3)$ \cite{Bottini:2025hri, Huang:2021ytb}.  Indeed, the SymTFT admits three topological boundary conditions associated with the Lagrangian algebras $\cL_{1,2,3}$ defined in \eqref{eq:L123}, on which are supported the $\H_{1,2,3}$ fusion categories, respectively.  

Now we may revisit the chiral $(\mathfrak{e}_8)_1$ WZW model in the SymTFT construction. The modular invariant partition function $\chi_0^{\cZ(\H_3)}   + \chi_{\pi_1}^{\cZ(\H_3)} + 2\, \chi_{\pi_2}^{\cZ(\H_3)}$ can be realized by choosing the topological boundary condition assciated with the Lagrangian algebra $\cL_3$. From the discussion in the previous paragraph, there is an $\H_3$ fusion category on the topological boundary, i.e. the $(\mathfrak{e}_8)_1$ model should have $\H_3$ symmetry.

Changing the Lagrangian algebra from $\cL_3$ to $\cL_1$ or $\cL_2$ leads to the third or second modular invariant partition function in (13) of the main text, i.e. $\chi_0^{\cZ(\H_3)}   + \chi_{\pi_1}^{\cZ(\H_3)} +  \chi_{\pi_2}^{\cZ(\H_3)}+\chi_{\sigma_1}^{\cZ(\H_3)}$ and $\chi_0^{\cZ(\H_3)}   + \chi_{\pi_1}^{\cZ(\H_3)} + 2\, \chi_{\sigma_1}^{\cZ(\H_3)}
$, respectively. In these cases, the topological boundary of the SymTFT supports  $\H_1$ or $\H_2$ fusion categories, and hence the $(\mathfrak{e}_8)_1$ model should have $\H_1$ and $\H_2$ symmetry as well.

The above result should not come as a surprise. It is known that $(\mathfrak{e}_8)_1$ is the only bosonic $c_-=8$ chiral CFT \cite{dong2002holomorphicvertexoperatoralgebras}, and the self-duality follows from this uniqueness.

The above discussion also shows that the $(\mathfrak{e}_8)_1$ model is self-dual under gauging any of the six algebras in Figure \ref{fig:gaugings2}. In particular,  half-space gauging of the algebras $1\oplus \alpha \oplus \alpha^2$ and $1\oplus \tilde{\alpha}\oplus \tilde{\alpha}^2$ give rise to conventional $\Z_3$ duality defects \cite{Thorngren:2019iar,Thorngren:2021yso,Kaidi:2021xfk,Kaidi:2022uux,Choi:2021kmx}, while half-space gauging of the remaining four algebra objects (all of which involve non-invertible lines) gives rise to more exotic duality defects \cite{Choi:2023vgk,Diatlyk:2023fwf,Perez-Lona:2023djo, Perez-Lona:2024sds,Yu:2025iqf, Seifnashri:2025fgd}. 
The full symmetry structure involving $\H_{1,2,3}$ and the six duality defects will be studied in upcoming work \cite{upcoming}.

\end{document}